\documentclass[fleqn]{elsart}
\usepackage{amsmath}
\usepackage{amstext}
\usepackage{amsfonts}
\usepackage{amssymb}
\usepackage{subfigure}
\usepackage{color}
\usepackage{epsfig}
\usepackage{bbm}

\newcommand{\eq}{\begin{equation}}
\newcommand{\en}{\end{equation}}

\journal{Nuclear Physics B}

\begin{document}

\begin{frontmatter}
  
  \vspace*{-1.0truecm} \title{
    $N_f = 2+1+1$ flavours of twisted mass quarks: cut-off effects at\\ tree-level of perturbation theory
  }
  \vspace*{-0.5truecm}

  \author[a,b]{E.V. Luschevskaya\corauthref{cor}},
  \ead{luschevskaya@gmail.com}
  \author[c]{Krzysztof Cichy},
  \ead{kcichy@amu.edu.pl}
  
  \corauth[cor]{Corresponding author.}
  
  \address[a]{ITEP, 117218 Russia, Moscow, B.\ Cheremushkinskaya str.\ 25}

  \address[b]{NIC, DESY, Platanenallee 6, D-15738 Zeuthen,
    Germany}
  
  \address[c]{Adam Mickiewicz University, Faculty of Physics\\
    Umultowska 85, 61-614 Pozna\'n, Poland}
    
  \begin{abstract}

\noindent We present a calculation of cut-off effects at tree-level of perturbation theory
for the K and D mesons using the twisted mass formulation of lattice 
QCD. The analytical calculations are performed in  the time-momentum frame.
The relative sizes of cut-off effects are compared for the pion, 
the kaon and the D meson masses. In addition, different realizations of maximal twist condition are considered
and the corresponding cut-off effects are analyzed.
    
  \end{abstract}

  \begin{keyword}
    Lattice gauge theory, twisted mass fermions, cut-off effects, 2+1+1 flavours of quarks
    \PACS 11.15.Ha, 12.38.Gc\\
    Preprint-No: DESY 10-245, ITEP-LAT/2010-14, SFB/CPP-10-131 \\
  \end{keyword}

\end{frontmatter}

\newpage

\section{Introduction}
Lattice calculations in QCD have shown significant advances in the last years 
\cite{Jansen:2008vs,Scholz:2009yz}. 
Simulations at the physical value of the pion mass are carried out
nowadays, including  up, down and strange quarks as 
dynamical degrees of freedom. 
A further step to render lattice QCD calculations even more realistic 
is also to include the charm quark in the simulations. 
In fact the simulations with two quark generations have already
been started \cite{Chiarappa:2006ae,Baron:2009zq,Baron:2010bv,Baron:2010th,Baron:2010vp,Bazavov:2010ru}.  
Clearly, having the dynamical charm 
degree of freedom allows to test many physical aspects of 
the charm sector of QCD, such as the mesonic and baryonic spectrum and decay
constants, heavy quark effects in operator matrix elements, 
the renormalized charm quark mass and eventually the running
of the strong coupling constant for four flavours. 

One concern when adding a charm quark mass is that lattice spacing effects 
may become large. Even in $\mathcal{O}(a)$-improved lattice theories 
cut-off effects of $\mathcal{O}(a^2m_\mathrm{charm}^2)$ are expected to 
be present which then can become significant due to the 
rather large value of the charm quark mass. 
In this paper, we want to investigate Wilson twisted mass fermions 
\cite{Frezzotti:2000nk,Frezzotti:2003xj,Frezzotti:2003ni}
in a formulation that comprises mass degenerate light up and down 
quarks and mass non-degenerate strange and charm quarks, which 
we  refer to as $N_f=2+1+1$ setup. 
All our calculations were performed at maximal twist, where 
automatic $\mathcal{O}(a)$-improvement is realized. 
The particular goal of this paper is to prepare the analytical 
basis for a study of the lattice spacing effects at tree-level of perturbation
theory. To this end, following Refs.~\cite{Carpenter:1984dd,Cichy:2008gk,Cichy:2008nt},
we derive the quark propagators both for the 4-dimensional 
discrete momentum representation and the time-momentum representation in this $N_f=2+1+1$ setup. From the 
expressions of these propagators we construct the full
(four times four) matrix of the correlation functions for K and D mesons.

Although calculations at tree-level of perturbation theory 
cannot be a quantitative measure for the interacting case when gluon degrees of freedom are incorporated, 
the qualitative results obtained in such setup can nevertheless serve as a valuable indicator. Studying  the  relative size of
lattice spacing effects for the pion, the K and D mesons  provides important information how cut-off effects behave
when light and heavy quarks are considered. 

In the twisted mass formulation of lattice QCD, the situation of maximal
twist is of particular importance since in this case all physical 
observables are automatically $\mathcal{O}(a)$-improved. We don't need an operator improvement and 
the calculation of corresponding coefficients. In the interacting case, maximal twist is realized 
by tuning of the Wilson quark mass to its critical value. As numerical and  theoretical investigations in the 
past have demonstrated, see
Refs.~\cite{Jansen:2005gf,Frezzotti:2005gi,Sharpe:2004ny,Aoki:2006nv,Jansen:2005kk}, 
there are, however, {\em optimal} and {\em non-optimal} ways to realize maximal twist. 

At tree-level of perturbation theory, we study the analogs of these definitions
of maximal twist. The optimal definition of maximal twist 
in the interacting case is to tune the PCAC quark mass to zero,
which corresponds to setting the bare Wilson quark mass
to zero at tree-level. The choice of tuning of the theory to maximal twist 
is not unique. Any definition of the critical quark mass that differs from 
the optimal one by $\mathcal{O}(a)$ effects still leads to maximal twist. However, such 
definitions can introduce unwanted, chirally enhanced $\mathcal{O}(a^2)$ effects, i.e. 
terms that go like $\mathcal{O}(a^2/m_\pi^2)$ and are hence called non-optimal. 
Our setup of tree-level of perturbation theory allows us to study both choices, 
the optimal tuning and the non-optimal tuning to maximal twist and we
explore both options in this paper to understand better the behaviour of meson masses 
as functions of $a^2$, when different tuning conditions are employed.

\section{Tree level of QCD with $N_f = 2+1+1$ flavours of quarks}


\subsection{Tree level of QCD in the continuum}


\subsubsection{Physical basis versus twisted basis}

We consider a light degenerate quark doublet 
$\psi^{(l)} = \{ \psi^{(u)} , \psi^{(d)} \}$ with a mass $m_{u,d}$ and a heavy 
non-degenerate quark doublet $\psi^{(h)} = \{ \psi^{(c)} , \psi^{(s)} \}$ with 
masses $m_c \neq m_s$. At tree-level of perturbation theory, i.e.\ in the 
absence of any gauge field, the light and heavy quark actions in the physical basis read:
\begin{equation}
 S_l = \int d^4x \, \bar{\psi}^{(l)}(x) \Big(\gamma_\mu \partial_\mu + m_{u,d}\Big) \psi^{(l)}(x), \label{EQN001}\\
\end{equation}
\begin{equation}
 S_h = \int d^4x \, \bar{\psi}^{(h)}(x) \Big(\gamma_\mu \partial_\mu + \textrm{diag}(m_c,m_s)\Big) \psi^{(h)}(x) \label{EQN002}.
\end{equation} 
 
Expressing the physical basis quark fields $\psi^{(l)}$ and $\psi^{(h)}$ in terms of 
twisted basis quark fields $\chi^{(l)}$ and $\chi^{(h)}$ via the twist rotation:
 \eq
 \psi^{(l)}=e^{i\omega_l \gamma_5 \tau_3/2} \chi^{(l)}, \ \ \ \psi^{(h)}=e^{i\omega_h \gamma_5 \tau_1/2} \chi^{(h)}
\label{EQN003} 
 \en
yields:
 \eq
 S_l = \int d^4x \, \bar{\chi}^{(l)}(x) \Big(\gamma_\mu \partial_\mu + m_{0,l} + i \mu_q \gamma_5 \tau_3\Big) \chi^{(l)}(x),
 \label{EQN004}
 \en
with: $m_{0,l} = m_{u,d} \cos(\omega_l)$, $\mu_q = m_{u,d} \sin(\omega_l)$ and
 \eq
 S_h = \int d^4x \, \bar{\chi}^{(h)}(x) \Big(\gamma_\mu \partial_\mu + m_{0,h} + i \mu_\sigma \gamma_5 \tau_1 + \mu_\delta \tau_3\Big) \chi^{(h)}(x),
 \label{EQN005}
 \en
where: $m_{0,h} = ((m_c + m_s)/2) \cos(\omega_h)$, $\mu_\sigma = ((m_c + m_s)/2) \sin(\omega_h)$ and 
$\mu_\delta = (m_c - m_s)/2$ respectively (the Dirac matrices 
$\gamma_\mu$ [$\mu = 0,1,2,3$] and $\gamma_5$ act in spinor space and the Pauli 
matrices $\tau_a$ [$a=1,2,3$] act in the flavour space), 
$\omega_l$ and $\omega_h$ are the so-called light and heavy twist angles.

The above relations between the quark masses and the mass parameters in the 
twisted basis quark actions can be solved with respect to the quark masses:
\begin{equation}
m_{u,d} = \sqrt{(m_{0,l})^2 + (\mu_q)^2} \label{EQN006}, \\
\end{equation}
\begin{equation}
m_s = \sqrt{(m_{0,h})^2 + (\mu_\sigma)^2} - \mu_\delta, \ \ \ m_c = \sqrt{(m_{0,h})^2 + (\mu_\sigma)^2} + \mu_\delta \label{EQN007}.
\end{equation}
At maximal twist $\omega_l = \omega_h = \pi/2$, corresponding to $m_{0,l} = m_{0,h} = 0$, these relations simplify to:
 \eq
 m_{u,d} = \mu_q, \ \ \ m_s = \mu_\sigma - \mu_\delta, \ \ \ m_c = \mu_\sigma + \mu_\delta .
 \label{EQN008}
 \en


\subsubsection{\label{SEC001}Meson spectrum}

At tree-level of QCD, i.e. in absence of gluonic fields and, therefore, of any 
interactions between quarks, mesons correspond to free quark-antiquark pairs. 
For mesons at rest, i.e. with total momentum $\mathbf{P} = 0$, the quark and the 
antiquark have to have opposite momenta $+\mathbf{p}$ and $-\mathbf{p}$. 
For example the spectra of K and D mesons are given by:
 \eq
 m_K(\mathbf{p}) = \sqrt{(m_s)^2 + \mathbf{p}^2} + \sqrt{(m_{u,d})^2 + \mathbf{p}^2}, \ \ \ m_D(\mathbf{p}) = \sqrt{(m_c)^2 + \mathbf{p}^2} + \sqrt{(m_{u,d})^2 + \mathbf{p}^2} .
 \label{EQN009}
 \en
When we consider infinitely extended space, there is a continuum of states 
parametrized by momentum $\pm \mathbf{p}$ associated with the quark-antiquark pair. 
When considering a finite spatial volume $L^3$, only discrete momenta 
$\mathbf{p} = 2 \pi \mathbf{n} / L$, $\mathbf{n} \in \mathbb{Z}^3$ are possible, 
rendering the meson spectra also discrete. In both cases all states are two-fold 
degenerate, corresponding to parity $\mathcal{P} = -$ and $\mathcal{P} = +$. 
An exception are mesons with quark momentum $\mathbf{p} = 0$, for which one can show 
that at tree-level they only exist for negative parity.


\subsection{Tree level of perturbation theory of twisted mass lattice QCD}

The lattice discretizations of the twisted basis quark actions (\ref{EQN004}) and (\ref{EQN005}) are: \cite{Frezzotti:2000nk,Frezzotti:2003xj}
\begin{equation}
S_l = a^4 \sum_x \bar{\chi}^{(l)}(x) \Big(D_\textrm{W}(m_{0,l}) + i \mu_q \gamma_5 \tau_3\Big) \chi^{(l)}(x), \label{EQN010} \\
\end{equation} 
\begin{equation}
S_h = a^4 \sum_x \bar{\chi}^{(h)}(x) \Big(D_\textrm{W}(m_{0,h}) + i \mu_\sigma \gamma_5 \tau_1 + \mu_\delta \tau_3\Big) \chi^{(h)}(x) \label{EQN011},
\end{equation} 
where $D_\mathrm{W}$ 
denotes the standard Wilson Dirac operator:
 \eq
 D_\mathrm{W}(m_{0,x}) = \frac{1}{2} \Big(\gamma_\mu \Big(\nabla_\mu + \nabla^\ast_\mu\Big) - a \nabla^\ast_\mu \nabla_\mu\Big) + m_{0,x}, \ \ \ x \in \{ l,h \} .\label{EQN012}
 \en

At maximal twist, physical observables are automatically $\mathcal{O}(a)$ 
improved, i.e.\ lattice discretization effects appear only 
quadratically \cite{Frezzotti:2003xj,Frezzotti:2003ni}.
Although maximal twist can be realized in many different ways, there is an 
optimal definition of maximal twist corresponding to setting 
$m_{0,l} = m_{0,h} = 0$ \cite{Frezzotti:2005gi}.
Note that for the light quark doublet, the Wilson term explicitly breaks isospin and 
parity, which becomes clear after a rotation to the physical basis. 
Only parity combined with light flavour exchange remains a symmetry. Since the Wilson 
term is $\mathcal{O}(a)$, isospin and parity are restored, when approaching the 
continuum limit. For the heavy quark doublet similar statements apply.

For infinite temporal extension, the lattice meson spectrum is expected to be 
qualitatively identical to the continuum meson spectrum discussed in Section~\ref{SEC001}. 
In particular, also on the lattice no positive parity mesons exist, with both 
quarks having zero momentum.


\section{Twisted mass lattice quark propagators}

The calculation of the heavy strange and charm quark propagators at tree-level of 
twisted mass lattice QCD is somewhat lengthy, but straightforward. Here we only 
quote the result, while details regarding the calculation can be found in App.\,A.

We can express the twisted mass action of the heavy doublet (\ref{EQN011}) in 
terms of the matrix $K(x;y)$:
\begin{eqnarray}
S_h = a^4 \sum_x \sum_y \bar{\chi}^{(h)}(x) K(x;y)  \chi^{(h)}(y) ,
\end{eqnarray}
which is of the form:
%
%
\begin{equation}
 K(x;y) = -\frac{1}{2 a} \sum_{\mu = 0}^3 \Big(\delta_{x+\hat{\mu},y} (1 - \gamma_\mu) + \delta_{x-\hat{\mu},y} (1 + \gamma_\mu)\Big)+ \delta_{x,y} \bigg(\Big(m_{0,h} + \frac{4}{a}\Big) + i \mu_\sigma \gamma_5 \tau_1 + \mu_\delta \tau_3\bigg),
\end{equation} 
where $x, y \in \mathbb{Z}$ denote space-time indices.
The equation:
\begin{eqnarray}
 a^4 \sum_y K(x;y) S^{(h)}(y;z) = \delta_{x,z}
\end{eqnarray}
relates $K(x;y)$ to the heavy twisted mass propagator $S^{(h)}(y;z)$ in position space representation.

In the time-momentum space representation, defined by:
\begin{eqnarray}
\mathbf{S}^{(h)}(t,\mathbf{p}) = \sum_{\mathbf{x}-\mathbf{y}} e^{-i \mathbf{p} (\mathbf{x}-\mathbf{y})} \mathbf{S}^{(h)}(x_0,\mathbf{x};y_0,\mathbf{y}) ,
\end{eqnarray}
where $t = x_0 - y_0$,
the resulting propagator for infinite temporal lattice extension reads:
\begin{eqnarray}
\mathbf{S}^{(h)}(\mathbf{p},\pm |t|) = A_{(1)} \bigg[\Big(N_{(1)} - i \mu_\sigma \gamma_5 \tau_1\Big) \Big(N_{(1)}^2 + R_{(1)}^2 - 2 \mu_\delta^2\Big) + \mu_{\delta} \tau_3 \Big(R_{(1)}^2 - N_{(1)}^2 - 2 \mu_\sigma^2\Big)\nonumber \\
- i \Big(N_{(1)}^2 + R_{(1)}^2 - 2 \mu_\sigma \mu_\delta \gamma_5 \tau_2 - 2 N_{(1)} \mu_\delta \tau_3\Big) \bigg({\mathcal K} \pm \frac{i \gamma_0}{a} \sinh E_{(1)}\bigg)\bigg] + \Big((1) \leftrightarrow (2)\Big), \label{EQN017}
\end{eqnarray}
with:
\begin{eqnarray}
\label{EQN018} & & A_{(1),(2)} = -\frac{a e^{-E_{(1),(2)} |t| / a}}{4 ((M_{0,h} + 1/a)^2 - \mu_\delta^2) \sinh E_{(1),(2)} (\cosh E_{(1),(2)} - \cosh E_{(2),(1)})}, \\
\label{EQN020} & & N_{(1),(2)} = M_{0,h} + \frac{1}{a} \Big(1 - \cosh E_{(1),(2)}\Big) , \ \ \ R_{(1),(2)}^2 = {\mathcal K}^2 + \mu_\sigma + \mu_\delta - \frac{1}{a^2} \sinh^2 E_{(1),(2)}, \\
\label{EQN013} & & {\mathcal K} = \frac{1}{a} \sum^3_{j=1} \gamma_j \sin(p_j a), \ \ \ M_{0,x} = m_{0,x} + \frac{2}{a} \sum^3_{j=1} \sin^2(p_j a / 2), \ \ \ x \in \{ l,h \},
\end{eqnarray}
and the poles of the propagator in the energy-momentum space representation:
\begin{eqnarray}
\label{EQN014} \cosh E_{(1),(2)} = \frac{-\tilde{b} \mp \sqrt{\tilde{b}^2 - 4 \tilde{a} \tilde{c}}}{2\tilde{a}} ,
\end{eqnarray}
where:
\begin{eqnarray}
 & & \tilde{a} = \frac{4}{a^2} \bigg(\bigg(M_{0,h} + \frac{1}{a}\bigg)^2 - \mu_\delta^2\bigg), \\
 & & \tilde{b} = \frac{8}{a} \mu_\delta^2 \bigg(M_{0,h} + \frac{1}{a}\bigg) - \frac{4}{a} s \bigg(M_{0,h} + \frac{1}{a}\bigg), \\
 & & \tilde{c} = s^2 - 4 \mu_\sigma^2 \mu_\delta^2 - 4 \mu_\delta^2 \bigg(M_{0,h} + \frac{1}{a}\bigg)^2, \\
 & & s = \bigg(M_{0,h} + \frac{1}{a}\bigg)^2 + \frac{1}{a^2} + {\mathcal K}^2 + \mu_\sigma^2 + \mu_\delta^2 .
\end{eqnarray}
This analytical result for the heavy propagator has been checked by comparing 
with numerically computed propagators for various spatial lattice extensions and quark masses.

The corresponding light twisted mass propagator has been calculated in \cite{Cichy:2008gk}. It reads:
\begin{eqnarray}
\label{EQN022} \mathbf{S}^{(l)}(\mathbf{p},\pm |t|) = B \Big(\Big(1 - \cosh E + a M_{0,l}\Big) \pm \gamma_0 \sinh E - i a {\mathcal K} - i a \mu_q \gamma_5 \tau_3\Big) ,
\end{eqnarray}
where ${\mathcal K}$ and $M_{0,l}$ are defined in (\ref{EQN013}) and
\begin{eqnarray}
\label{EQN023} B = \frac{e^{-E |t| / a}}{2 \sinh E (1 + a M_{0,l})} , \ \ \ \cosh E = \frac{(a M_{0,l} + 1)^2 + {\mathcal K}^2 a^2 + \mu_q^2 a^2 + 1}{2 (1 + a M_{0,l})} .
\end{eqnarray}


\section{Correlation matrices for K and D mesons}
\label{sec:four}
To study cut-off effects at tree-level of perturbation theory for $N_f = 2+1+1$ quark 
flavours, we consider the spectrum of K and D mesons. As already discussed 
in Section~\ref{SEC001}, such mesons consist of the light up/down antiquark and 
either the heavy strange or the heavy charm quark. Since there is no gluonic field at 
tree-level, both quarks are free particles.

To create such mesons with well defined quantum numbers, we apply meson creation 
operators $\mathcal{O}_{(h,\Gamma)}$ to the vacuum state $| \Omega \rangle$. 
In the physical basis in continuum QCD, a possible choice of appropriate operators is given by:
\begin{eqnarray}
\mathcal{O}_{(h,\Gamma)}(t) = \int d^3x \, \bar{\psi}^{(u)}(\mathbf{x},t) \Gamma \psi^{(h)}(\mathbf{x},t) .
\end{eqnarray}
The heavy flavour index $h \in \{ s , c \}$ determines whether the K meson or the
D meson is created, the $4 \times 4$ matrix $\Gamma \in \{ \gamma_5 , 1 \}$ realizes the
total angular momentum $J = 0$ and either the negative parity $\mathcal{P} = -$ or the
positive parity $\mathcal{P} = +$ and the integration over space $\int d^3x$ yields the 
total momentum $\mathbf{P} = 0$, i.e.\ assures that the light and the heavy quark 
have opposite momenta $\pm \mathbf{p}$. In principle, one could also fix the 
individual quark momenta $\pm \mathbf{p}$ by including a second integration 
over space, but we prefer to consider not only the ground state, but also higher states, 
to have a situation, which is more like one of the interacting case, 
i.e.\ beyond tree-level, where one cannot get rid of excited states at the 
stage of operator construction, see Refs.~\cite{Baron:2010th,Baron:2010vp} for an
investigation of meson mass 
determinations in the interacting case.

On the lattice using the twisted mass formalism an equivalent set of meson 
creation operators is given by:
\begin{eqnarray}
\label{EQN669} \mathcal{O}_{(h,\Gamma)}(t) = \sum_\mathbf{x} \, \bar{\chi}^{(u)}(\mathbf{x},t) \Gamma \chi^{(h)}(\mathbf{x},t) .
\end{eqnarray}
Note, however, that at finite lattice spacing, even after a rotation to the physical 
basis, parity and heavy flavour are only approximate quantum numbers, because of 
explicit twisted mass flavour and parity breaking. For a detailed discussion of 
these issues we refer to \cite{Baron:2010th,Baron:2010vp}. To extract meson masses, we first calculate $4 \times 4$ correlation matrices 
with the four operators (\ref{EQN669}):
\eq
C_{(h_1,\Gamma_1),(h_2,\Gamma_2)}(t_1-t_2) = \langle \Omega | \Big(\mathcal{O}_{(h_1,\Gamma_1)}(t_1)\Big)^\dagger \mathcal{O}_{(h_2,\Gamma_2)}(t_2) | \Omega \rangle .
\en
Details of this calculation, which uses the propagators from the previous section, 
are presented in App.\,B. Then, we extract the masses of the K and D mesons, as explained in details in App.\,C and \cite{Baron:2010th}.


\section{Numerical results}

In this section, we will present some results for the continuum limit scaling 
of meson masses using maximally twisted mass fermions. In the investigation below, we will 
employ both the optimal and non-optimal definitions of maximal twist. 


\subsection{Setup and physical parameters}

We consider a setup, which is reminiscent of QCD with 
$N_f = 2+1+1$ flavours of quarks: there is the degenerate doublet of light fermions 
(masses $m_{u,d}$) and the  non-degenerate doublet of significantly heavier fermions 
(masses $m_s \approx 22 m_{u,d}$ and $m_c \approx 284 m_{u,d}$), 
where the ratios of masses have been chosen as observed in the nature for the up/down, 
strange and charm quarks.
However, in contrast to QCD, we consider the fermions (to which we will also 
refer as ``quarks'') at tree-level, i.e.\ there are no interactions of any kind.

We consider a $3$-dimensional spatial volume of extension $N^3$ with periodic boundary conditions.
Having fixed the ratios of quark masses, there are two dimensionful parameters 
remaining, $N$ and $m_{u,d}$. Their dimensionless product $N m_{u,d}$ fully determines 
the physical situation. We choose $N m_{u,d} = 0.01$. Roughly speaking, this 
assures that even for our smallest lattices, the heavy charm quark mass 
$m_c$ in lattice units is smaller than $1$ (see below for more details).


\subsection{Continuum limit and lattice parameters}

For our lattice computations, the following set of parameters has to be chosen: 
the number of lattice sites $N$ in spatial direction 
(i.e.\ the spatial lattice volume is $N^3$), the twisted quark masses 
$\mu_q$, $\mu_\sigma$, $\mu_\delta$ (in units of $N=1/a$) and the untwisted 
quark masses $m_{0,l}$ and $m_{0,h}$ (in units of $N=1/a$).

At tree-level of perturbation theory, the continuum limit ($a\rightarrow0$) and the infinite volume limit ($N\rightarrow\infty$) are equivalent.  This means that the role of lattice spacing $a$ is played simply by the inverse of the number of lattice points $N$. 

To recover in the continuum limit the setup described 
in the previous subsection, the lattice parameters have to be chosen in the following way,
if the optimal definition of maximal twist is used:
\begin{itemize}
\item Untwisted quark masses:
\begin{eqnarray}
m_{0,l} = m_{0,h} = 0 .
\end{eqnarray}

\item Twisted quark masses:
\begin{eqnarray}
\label{EQN569} N \mu_q =  m_{u,d}, \ \ \ N(\mu_\sigma - \mu_\delta) = m_s , \ \ \ N(\mu_\sigma + \mu_\delta) = m_c.
\end{eqnarray}
\end{itemize}

When approaching the continuum limit, we set the twisted quark masses as 
specified in (\ref{EQN569}), i.e.\ we use this choice of twisted quark masses 
also at finite lattice spacing. 
The actual values of the strange and the charm quark mass for our numerical simulations are chosen such that they correspond to the ratios of central values of quark mass estimates published by the Particle Data Group \cite{PDG} -- $m_s/m_{u,d}=21.5901$ and $m_c/m_{u,d}=284.091$\footnote{Of course, since we are studying here an unphysical situation which, however, can illustrate the size of 
cut-off effects, the definite values for the quark masses used are not 
very important and any reasonable value of the ratio of quark masses would 
be sufficient. Nevertheless, we have used the accurate values of ref. \cite{PDG} to 
reflect the physical ratios of the quark masses as close as possible.}.
This implies that since 
$N (\mu_\sigma + \mu_\delta) = N m_c \approx 0.01 \times 284 = 2.84$, even for 
our smallest lattices, corresponding to $N = 4$, the heaviest quark mass in 
lattice units (equal to $\mu_\sigma + \mu_\delta$) is smaller than $1$.

Regarding the untwisted quark masses, we study different ways of approaching the 
continuum limit. A particular and optimal choice is $m_{0,l} = m_{0,h} = 0$.
However, any other choice $N m_{0,l} = \mathcal{O}(1/N)$ and $N m_{0,h} = \mathcal{O}(1/N)$ 
also corresponds to maximal twist. Therefore, we also study the choices:
\begin{eqnarray}
N m_{0,l} = \frac{c_l}{N} , \ \ \ N m_{0,h} = \frac{c_h}{N},
\end{eqnarray}
with $c_l$ and $c_h$ arbitrary, but constant.
Note, however, that in practice such choices should not be considered since 
--while still keeping the $\mathcal{O}(a)$-improvement of the theory-- they 
might lead to the chirally enhanced cut-off effects, as discussed in
Refs.~\cite{Jansen:2005gf,Frezzotti:2005gi,Sharpe:2004ny,Aoki:2006nv,Jansen:2005kk}.


\subsection{Numerical study of the continuum limit}

\begin{figure}[t]
\centering
\subfigure[\label{fig:massesKsmallc}]{\includegraphics[height=3.1in, angle=-90]{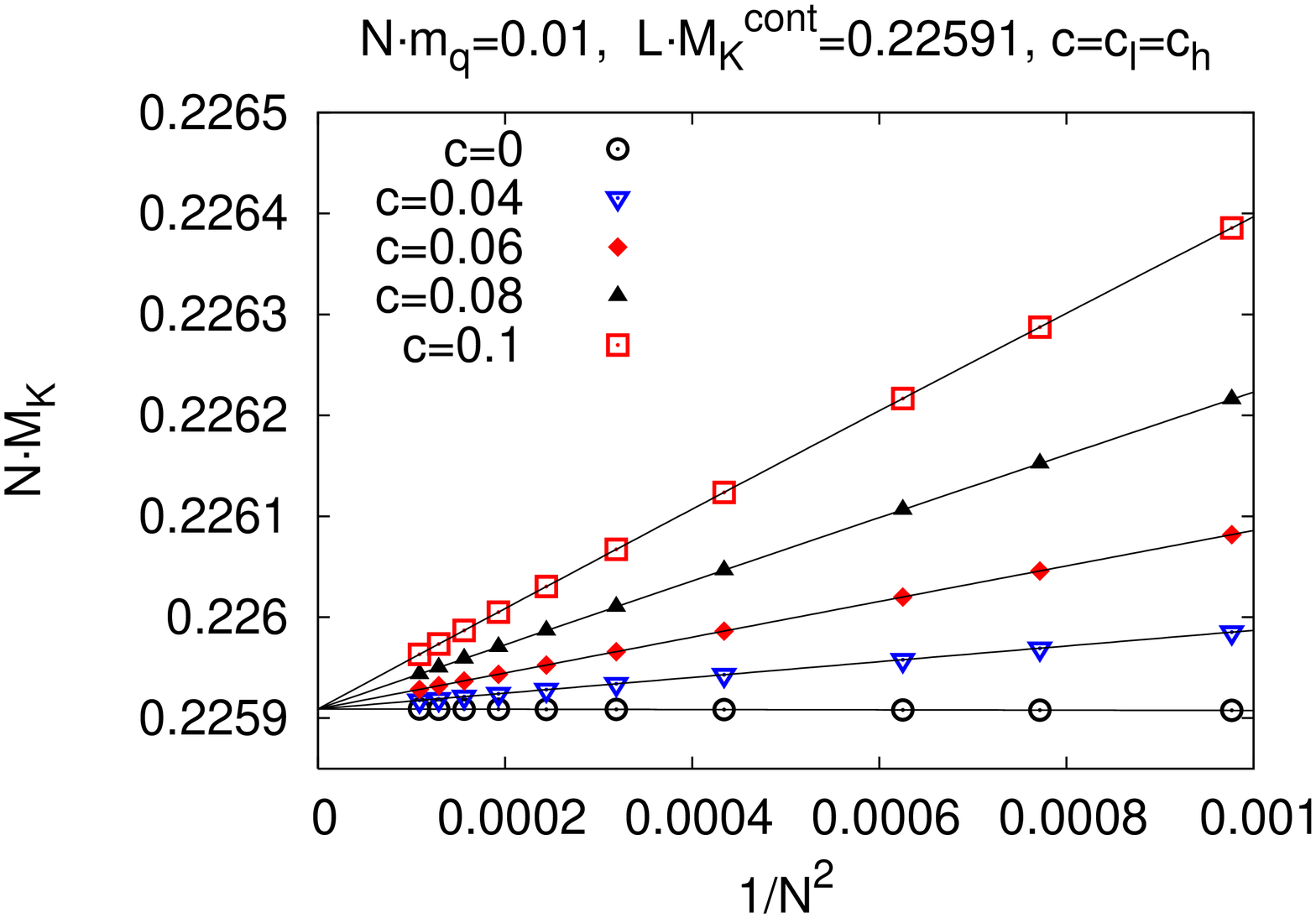}}
\subfigure[\label{fig:massesDsmallc}]{\includegraphics[height=3.1in, angle=-90]{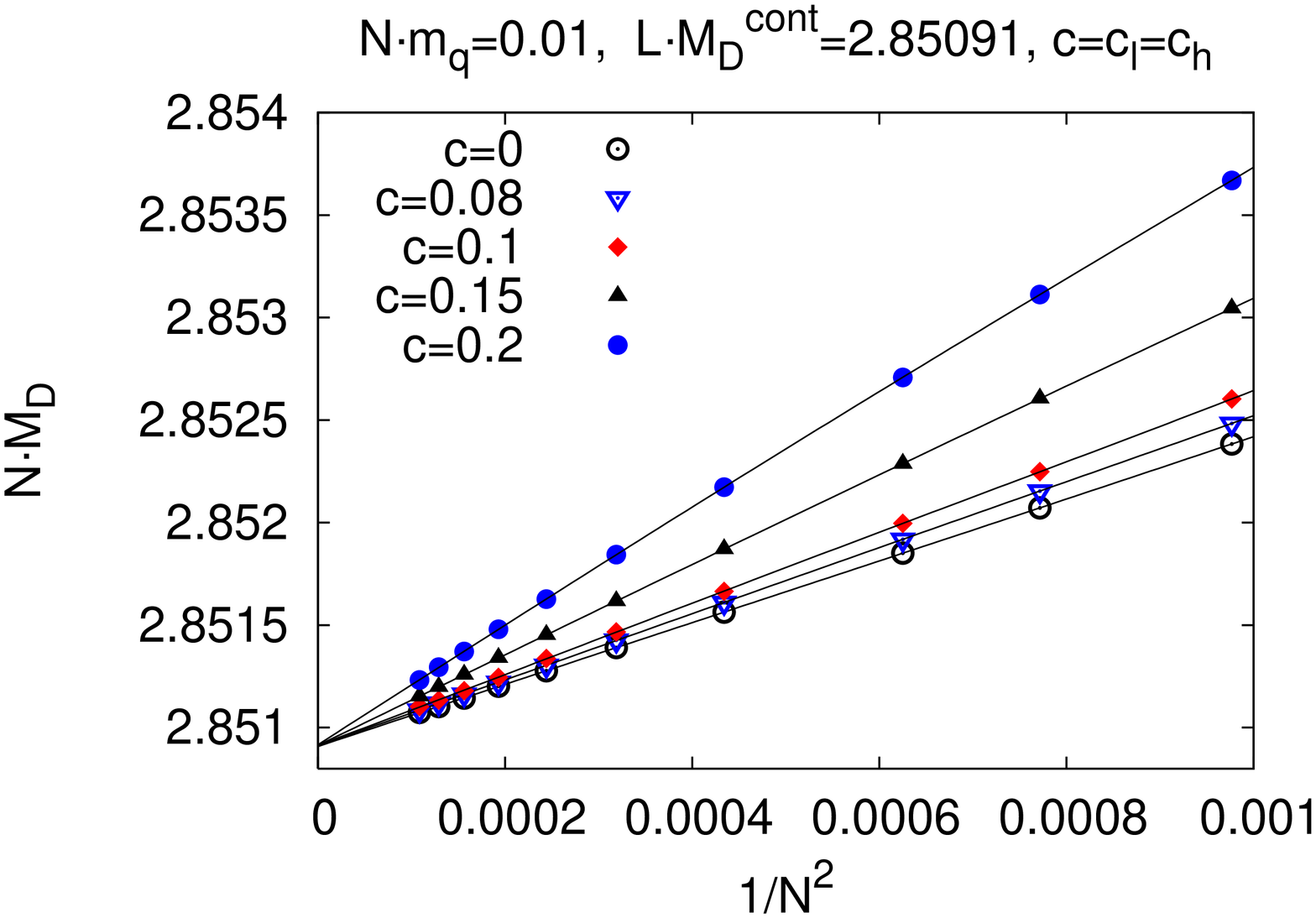}}
\caption{\label{fig:massesKDsmallc} The cut-off effects and the continuum limit of: (a) kaon mass, (b) D meson mass, for different realizations of maximal twist $c\equiv c_l = c_h \in \{0, 0.04, 0.06, 0.08, 0.1 \}$ and $32 \leq N \leq 96$ lattices.}
\end{figure}

\begin{figure}[t]
\centering
\subfigure[\label{fig:massesKlargec}]{\includegraphics[height=3.1in, angle=-90]{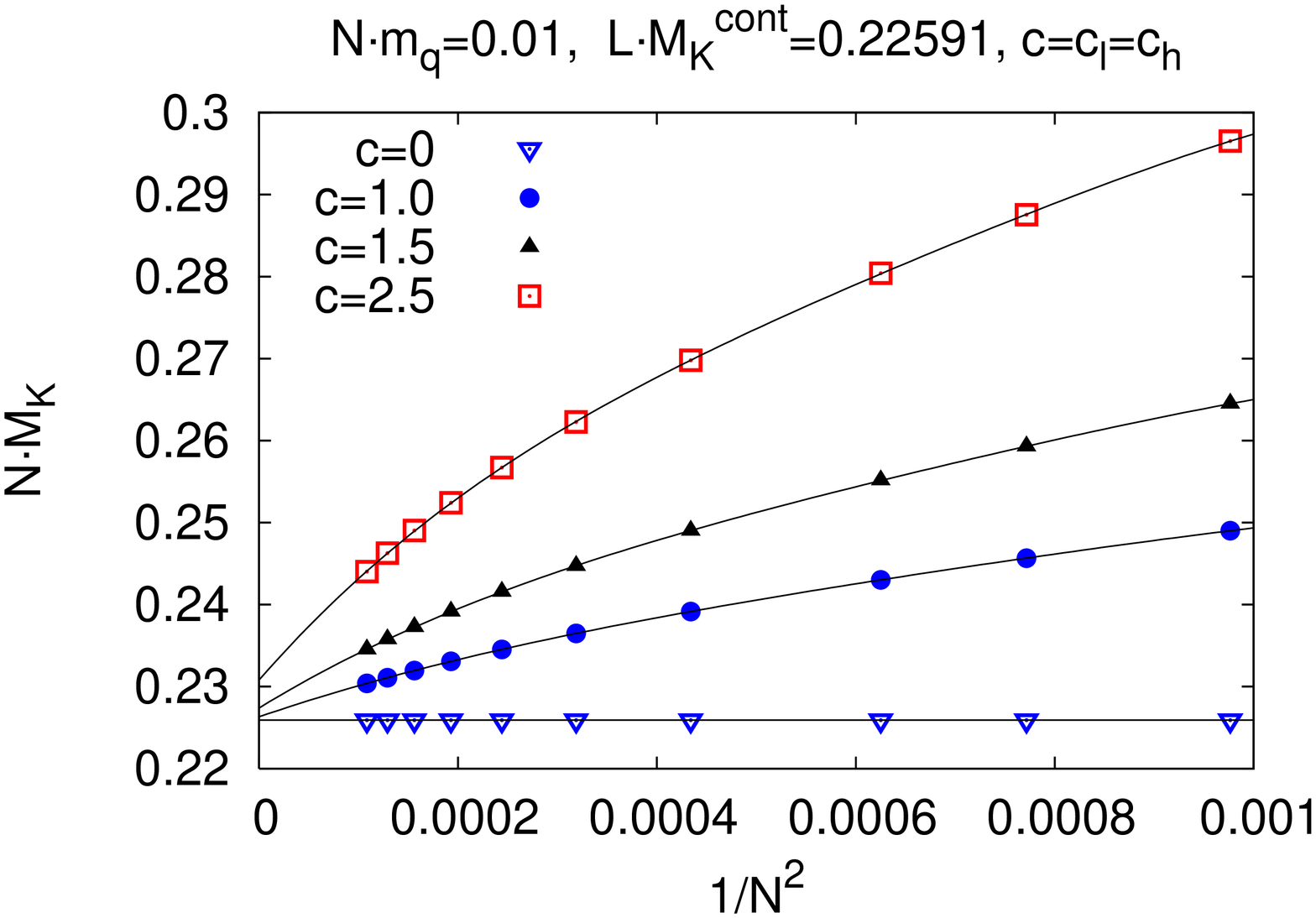}}
\subfigure[\label{fig:massesDlargec}]{\includegraphics[height=3.1in, angle=-90]{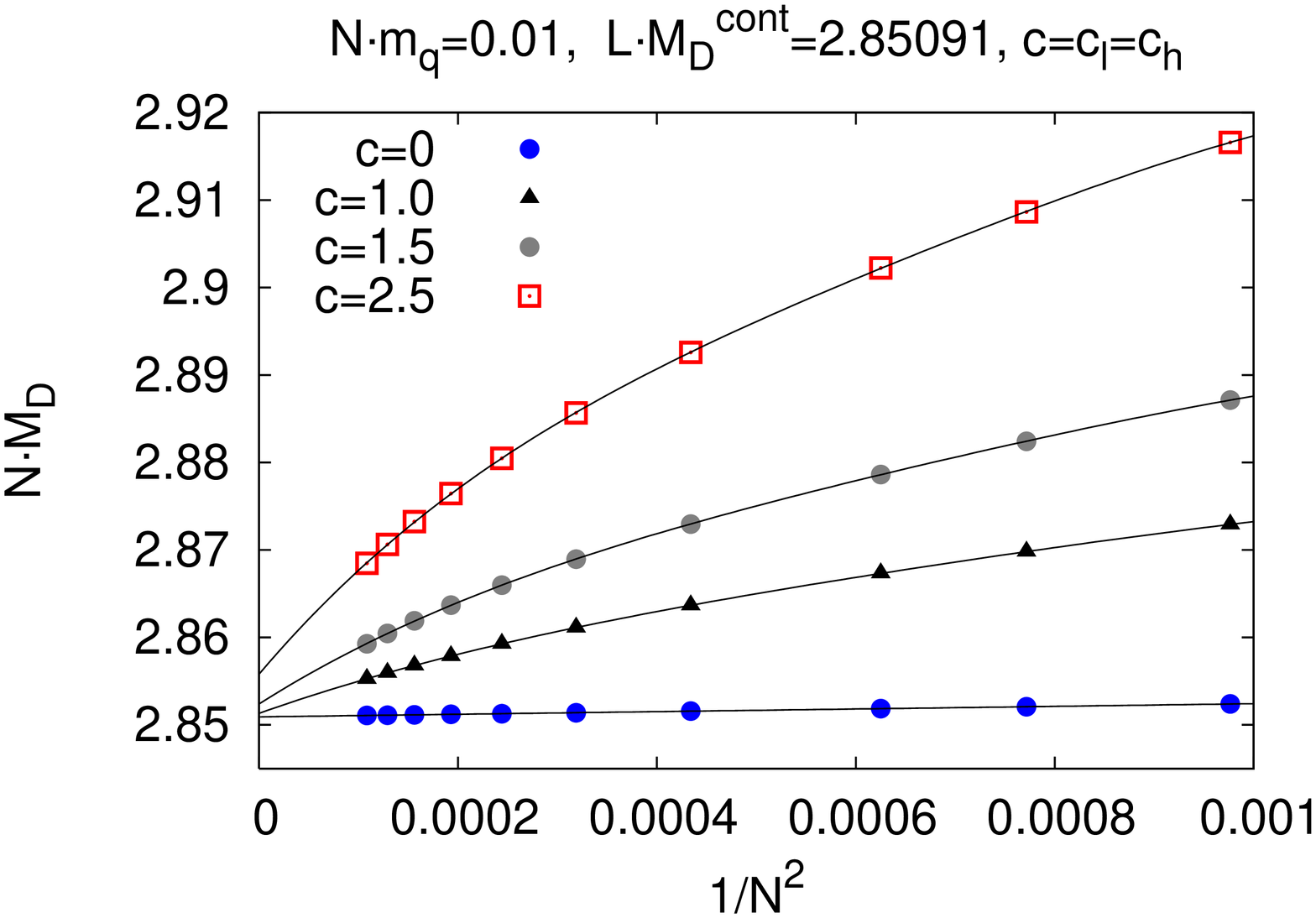}}
\caption{\label{fig:massesKDlargec} The cut-off effects and the continuum limit of: (a) kaon mass, (b) D meson mass, for different realizations of maximal twist $c\equiv c_l = c_h \in \{0, 1.0, 1.5, 2.5\}$ and $32 \leq N \leq 96$ lattices.}
\end{figure}

\begin{figure}[t]
\centering
\subfigure[\label{fig:massesKdifferentc}]{\includegraphics[height=3.1in, angle=-90]{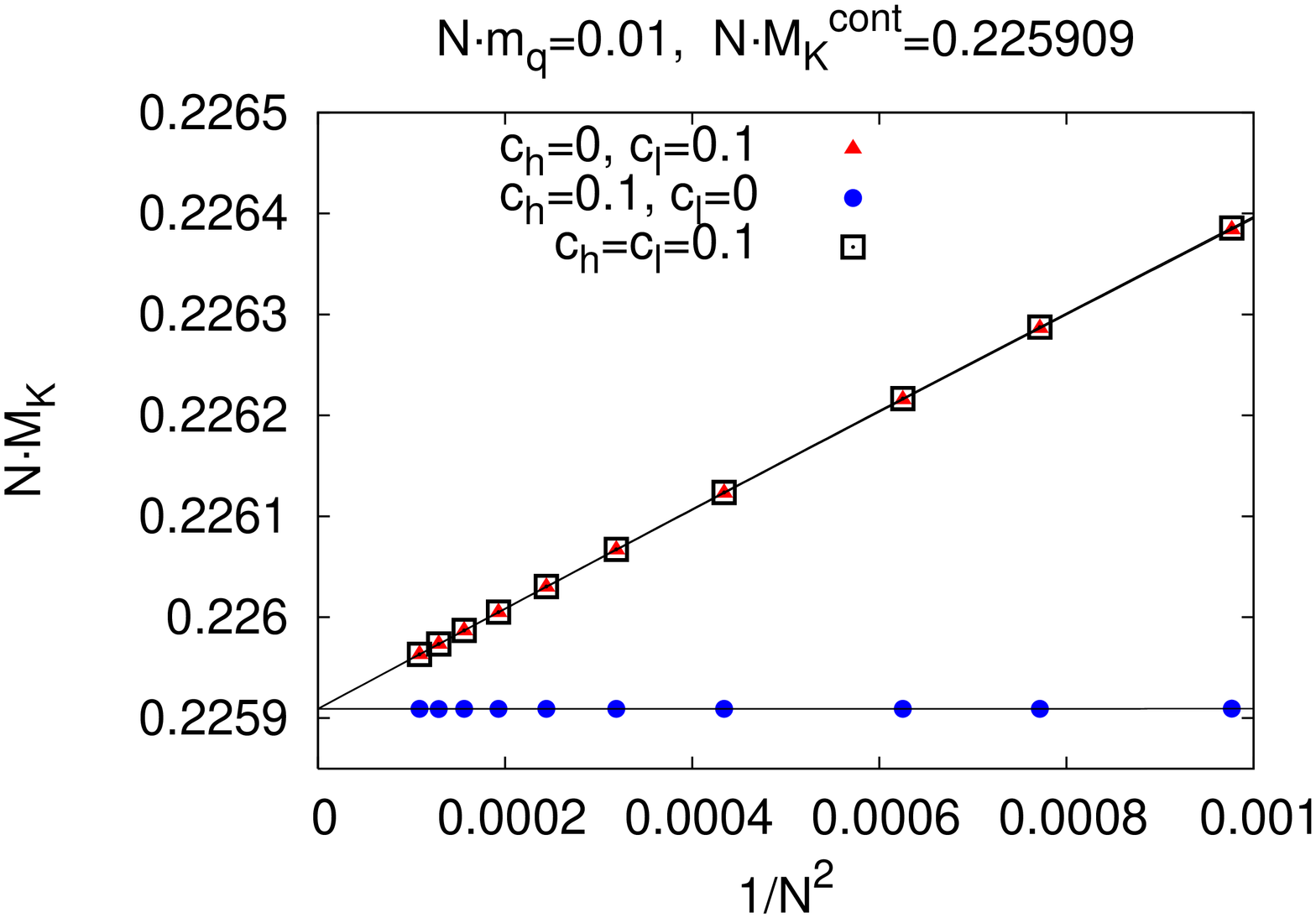}}
\subfigure[\label{fig:massesDdifferentc}]{\includegraphics[height=3.1in, angle=-90]{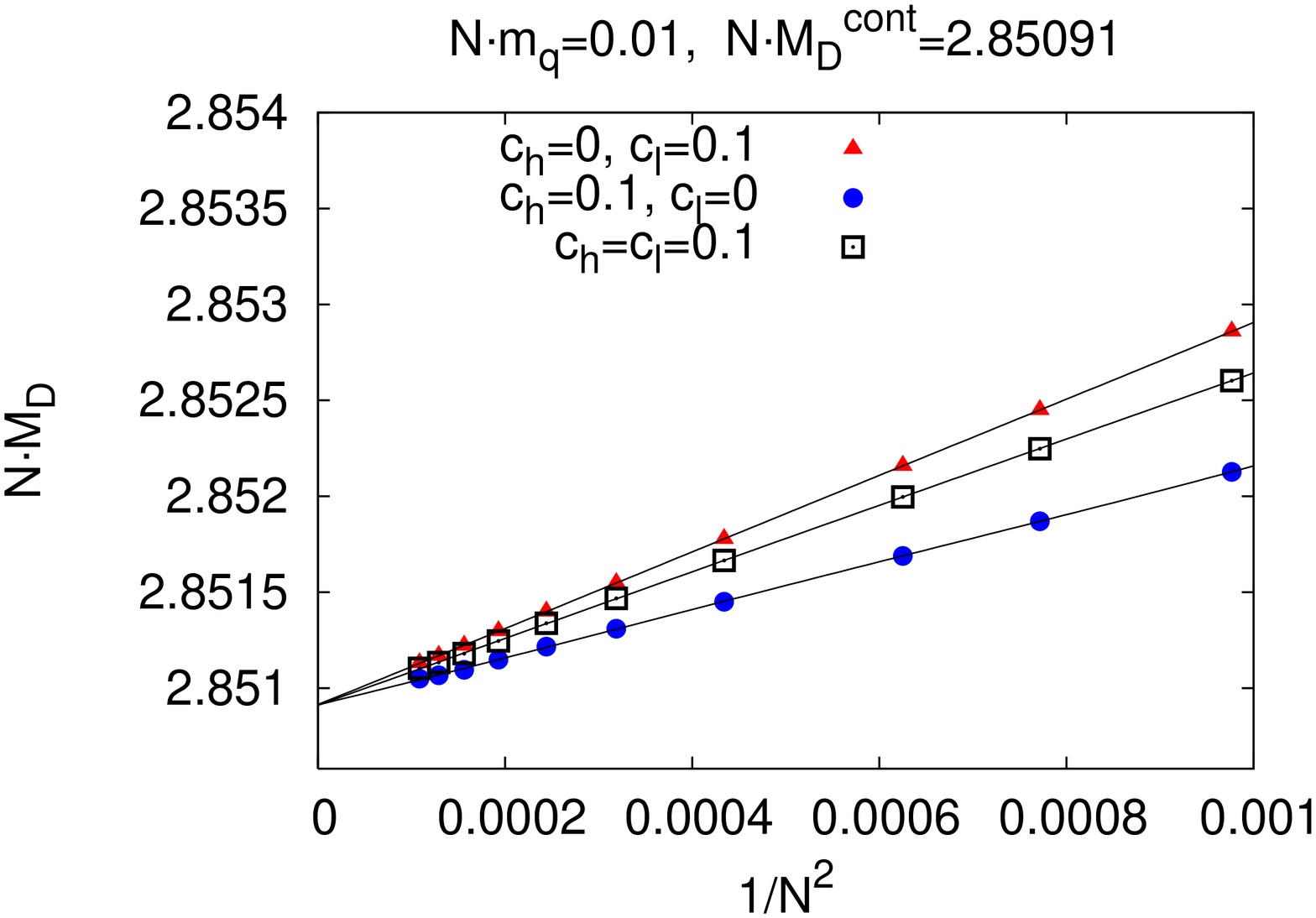}}
\caption{\label{fig:massesKDdifferentc} The cut-off effects and the continuum limit of: (a) kaon mass, (b) D meson mass, for different realizations of maximal twist: $(c_l=0,\,c_h\neq0)$, $(c_l\neq0,\,c_h=0)$, $(c_l=c_h\neq0)$.}
\end{figure}

We study the continuum limit by performing computations with various lattice 
volumes, ranging from $N^3 = 4^3$ to $N^3 = 96^3$.
In Figs.~\ref{fig:massesKDsmallc}, \ref{fig:massesKDlargec} and \ref{fig:massesKDdifferentc} we investigate the continuum limit of $m_K$ and $m_D$ 
for different realizations of maximal twist, $c\equiv c_l = c_h \in \{0, 0.04, 0.06, 0.08, 0.1 \}$ (Fig.~\ref{fig:massesKDsmallc}), $c\equiv c_l = c_h \in \{0, 1.0, 1.5, 2.5\}$ (Fig.~\ref{fig:massesKDlargec}) and $c_l=0, c_h=0.1$ or $c_l=0.1, c_h=0$ or $c_l=c_h=0.1$ (Fig. \ref{fig:massesKDdifferentc}).
The curves in the plots correspond to fits of quartic polynomials in $1 / N^2$:
\begin{equation}
\label{EQN611} N m_{\pi,K,D} = a_0 + a_1 \frac{1}{N^2} + a_2 \left(\frac{1}{N^2}\right)^2 + a_3 \left(\frac{1}{N^2}\right)^3 + a_4 \left(\frac{1}{N^2}\right)^4.
\end{equation} 

In the former case, i.e. for relatively small values of the parameter $c$ and large lattices ($N=32$ to $N=96$), we observe a linear dependence in $1/N^2 = a^2$ (i.e. the values of higher order coefficients $a_n$ ($n\geq2$) in (\ref{EQN611}) are very small). This is, of course, expected, since Wilson 
twisted mass lattice fermions at maximal twist guarantee the absence of $\mathcal{O}(a)$ discretization effects.


On the other hand, for large values of 
$c\equiv c_l = c_h$ (as shown in Fig. \ref{fig:massesKDlargec}),
both K and D meson masses as functions of $1 / N^2$ clearly 
exhibit a non-vanishing curvature increasing with the value of $c$. 
However, this curvature can be well described by higher order corrections
in $1/N^2$, as required from an $\mathcal{O}(a)$-improved theory.
Nevertheless, when choosing the coefficients $c_{l,h}$ large, the theory is more and more
tuned towards the Wilson quark action showing large cut-off effects.
For very large $c$ values, also the extracted continuum limit values
of meson masses are not reliable -- because of the large curvature, in order to extrapolate
to the continuum limit, one would need lattices with $N>96$ in such case.
Of course, the values of $c_{l,h}$ used here are exceptionally large, leading
to cases of non-optimal tuning to maximal twist that would never be used 
in practice. Hence, our investigation of these non-optimal tuning conditions
serves solely illustrative purposes to indicate when higher order corrections 
in $1/N^2$ become relevant.

%
%
%
%
%
%
%


\begin{figure}[t]
\centering
\subfigure[\label{fig:slope1}]{\includegraphics[height=3.1in, angle=-90]{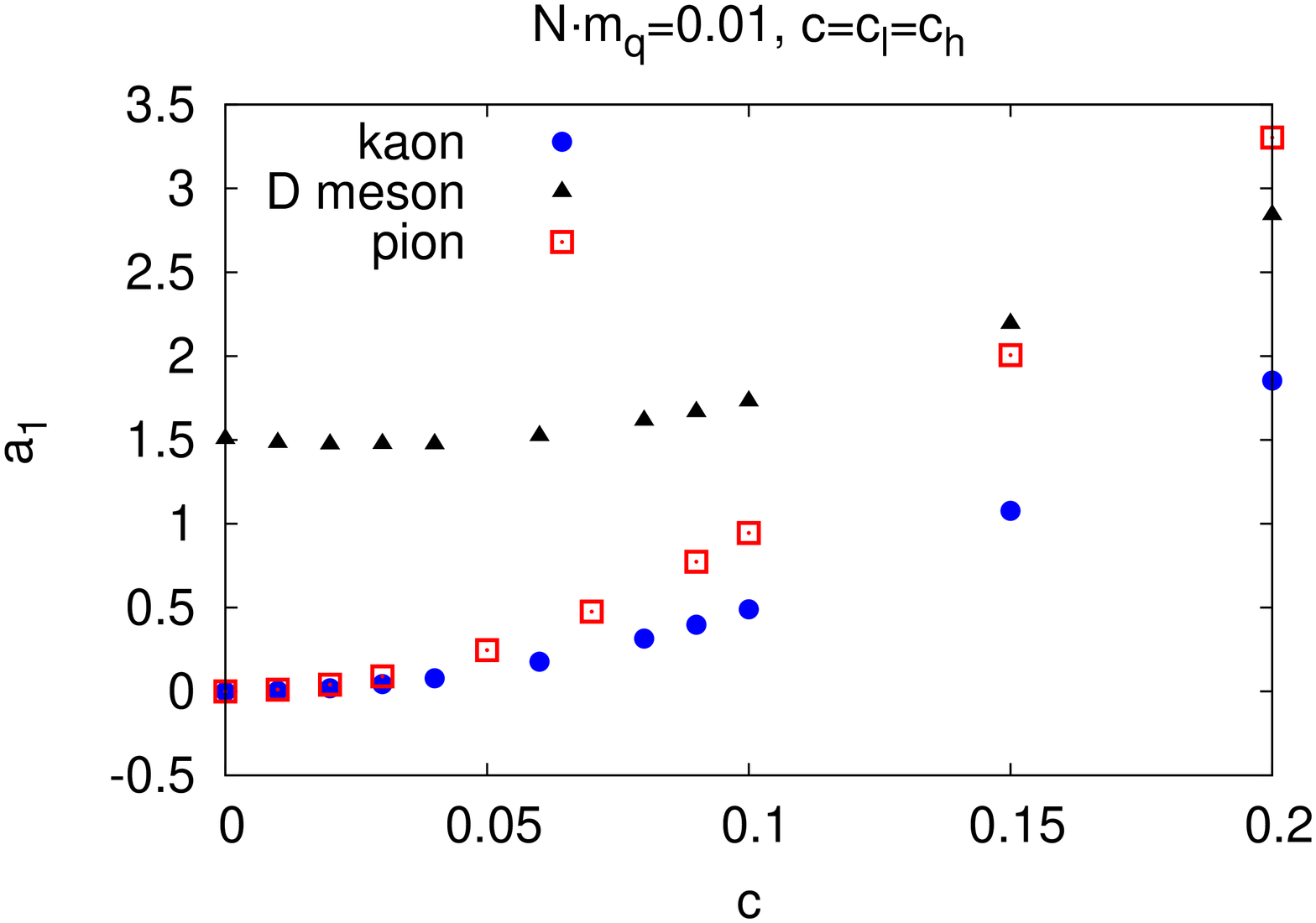}}
\subfigure[\label{fig:slope2}]{\includegraphics[height=3.1in, angle=-90]{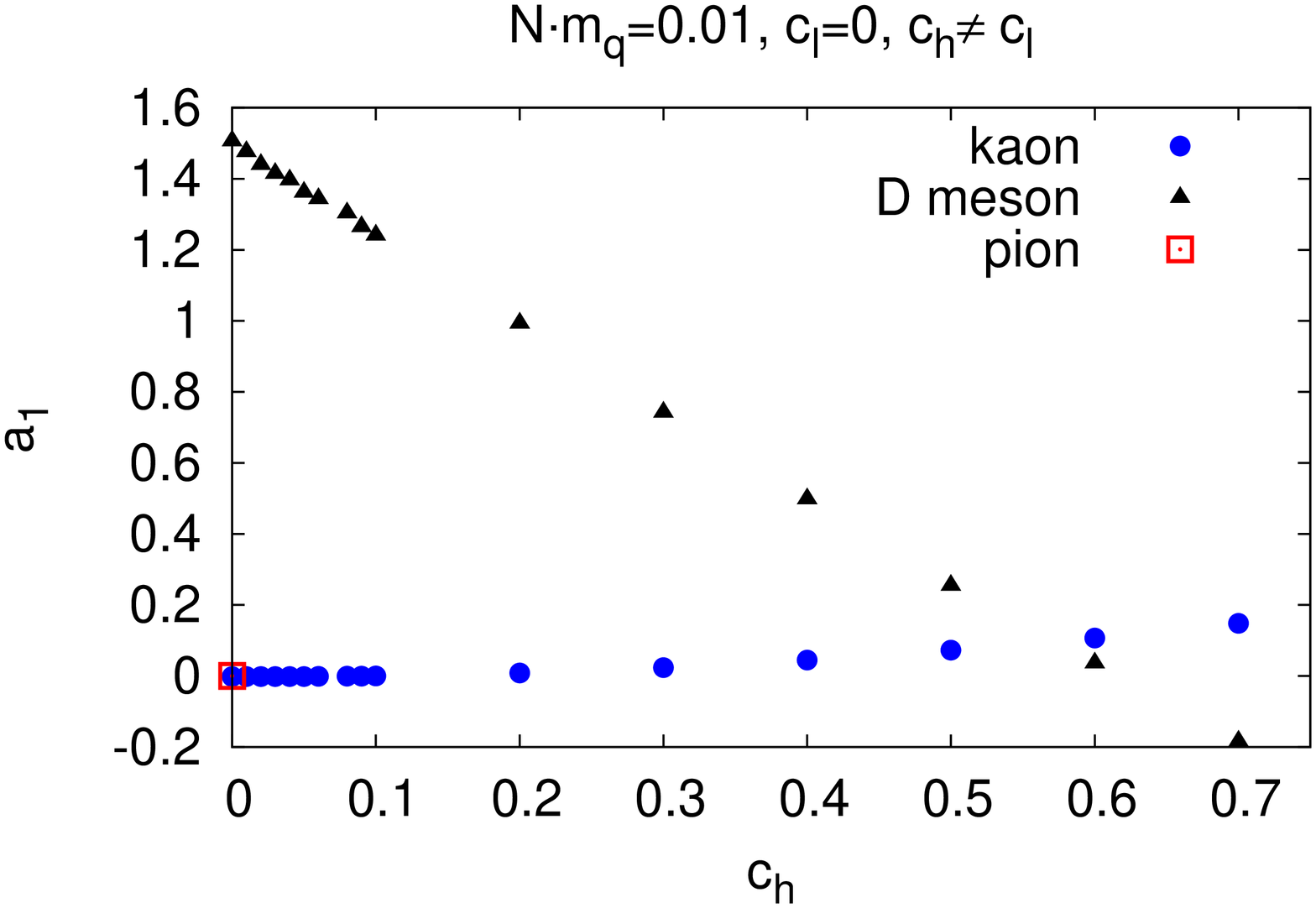}}
\caption{\label{fig:slope} The slope $a_1$ in eq.~(\ref{EQN611}) for the pion, kaon and D meson masses vs. the values of $c_l$ and $c_h$. (a) The case $c\equiv c_l = c_h$. (b) The case $c_l=0,\,c_h\geq0$.}
\end{figure}

The values obtained for the $a_0$ coefficient correspond to meson masses extrapolated 
to the continuum limit. Within our numerical precision, they agree with the expected 
continuum results, i.e. the respective sums of the corresponding quark masses 
$N m_{u,d} = 0.010$, $N m_s = 0.21591$ and $N m_c = 2.84091$.

The slope parameter $a_1$ describes the magnitude of $\mathcal{O}(a^2)$ discretization effects.
The extracted values of this parameter are shown in Fig.~\ref{fig:slope}
for the cases $c\equiv c_l = c_h$ and $c_l=0,\,c_h\geq0$.
In the case of $c_l=c_h=0$ (Fig.~\ref{fig:slope1}), the cut-off effects are significantly larger
for the D meson than for the K meson, while the smallest ones are observed for the pion.
Considering the relative discretization effects, given by the ratios $a_1 / a_0$,
we observe that the cut-off effects are $\sim10^2$ times larger for the D meson than for the K meson
and $\sim10^3$ times larger for the kaon than for the pion.
Moreover, the size of discretization effects increases with increasing values of the parameter $c\equiv c_l = c_h$, but the sensitivity to this parameter is clearly the largest for the pion mass and it is very weak in the case of the D meson mass. This is an expected feature of the considered setup, since the non-vanishing bare Wilson quark mass, introduced by non-zero values of the parameter $c$, is relatively large in comparison with the pion mass and almost negligible as compared to the D meson mass (unless very large values of $c$ are considered). 
The case $c_l=0,\,c_h\geq0$ (Fig.~\ref{fig:slope2}) will be referred to below.

However, as we have already mentioned, $\mathcal{O}(a^2m_q^2)$ effects are expected to be present in $\mathcal{O}(a)$-improved theories and they can become important
in theories with heavy strange and charm quarks. Therefore, the $\mathcal{O}(a^2)$ discretization effects
given by the extracted parameter $a_1$ contain the ``pure'' $\mathcal{O}(a^2)$ effects and in addition the $\mathcal{O}(a^2m_q^2)$ effects.

\begin{figure}[t]
\centering
\subfigure[\label{fig:Nms_effects}]{\includegraphics[height=3.1in, angle=-90]{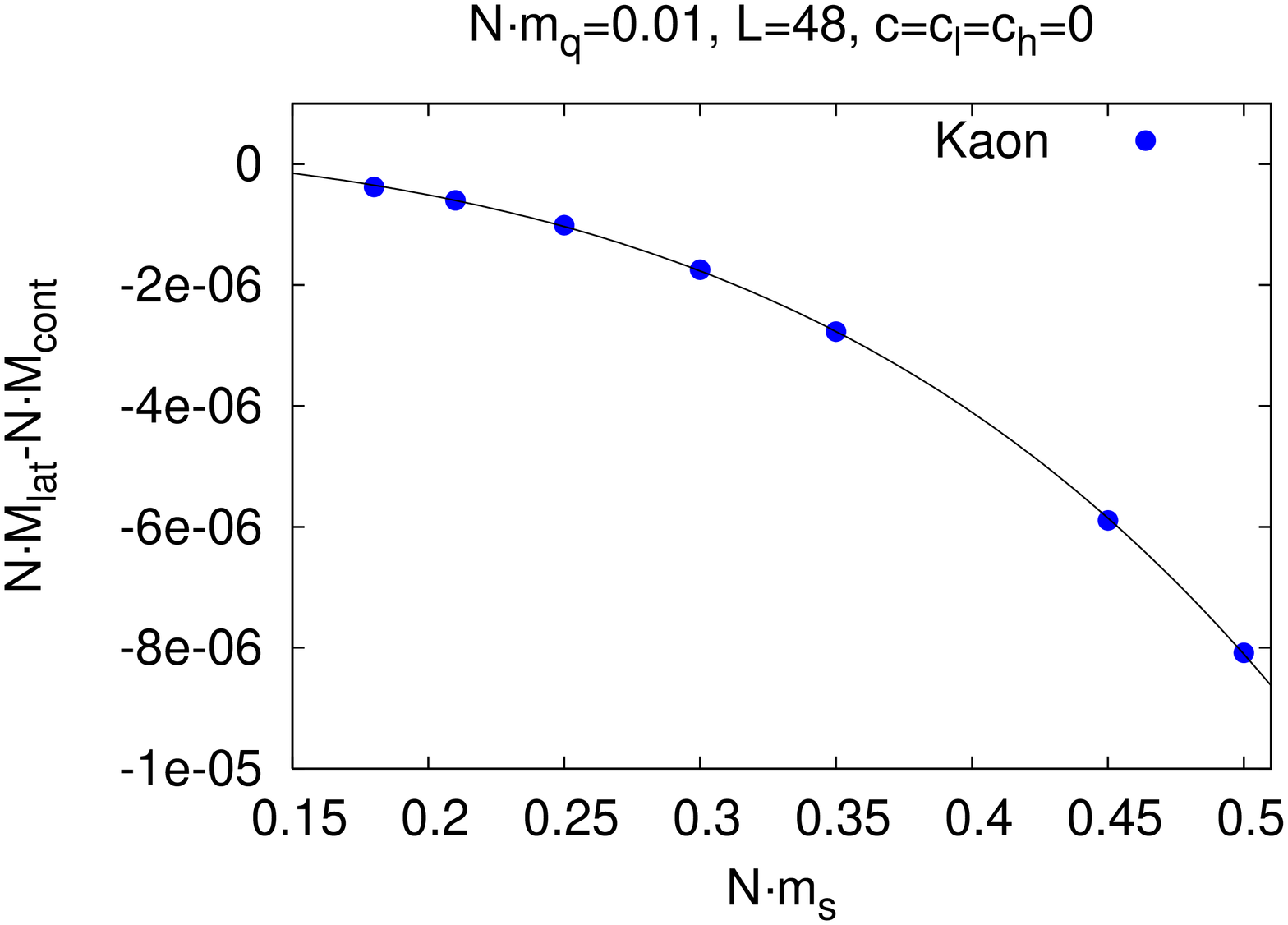}}
\subfigure[\label{fig:Nmc_effects}]{\includegraphics[height=3.1in, angle=-90]{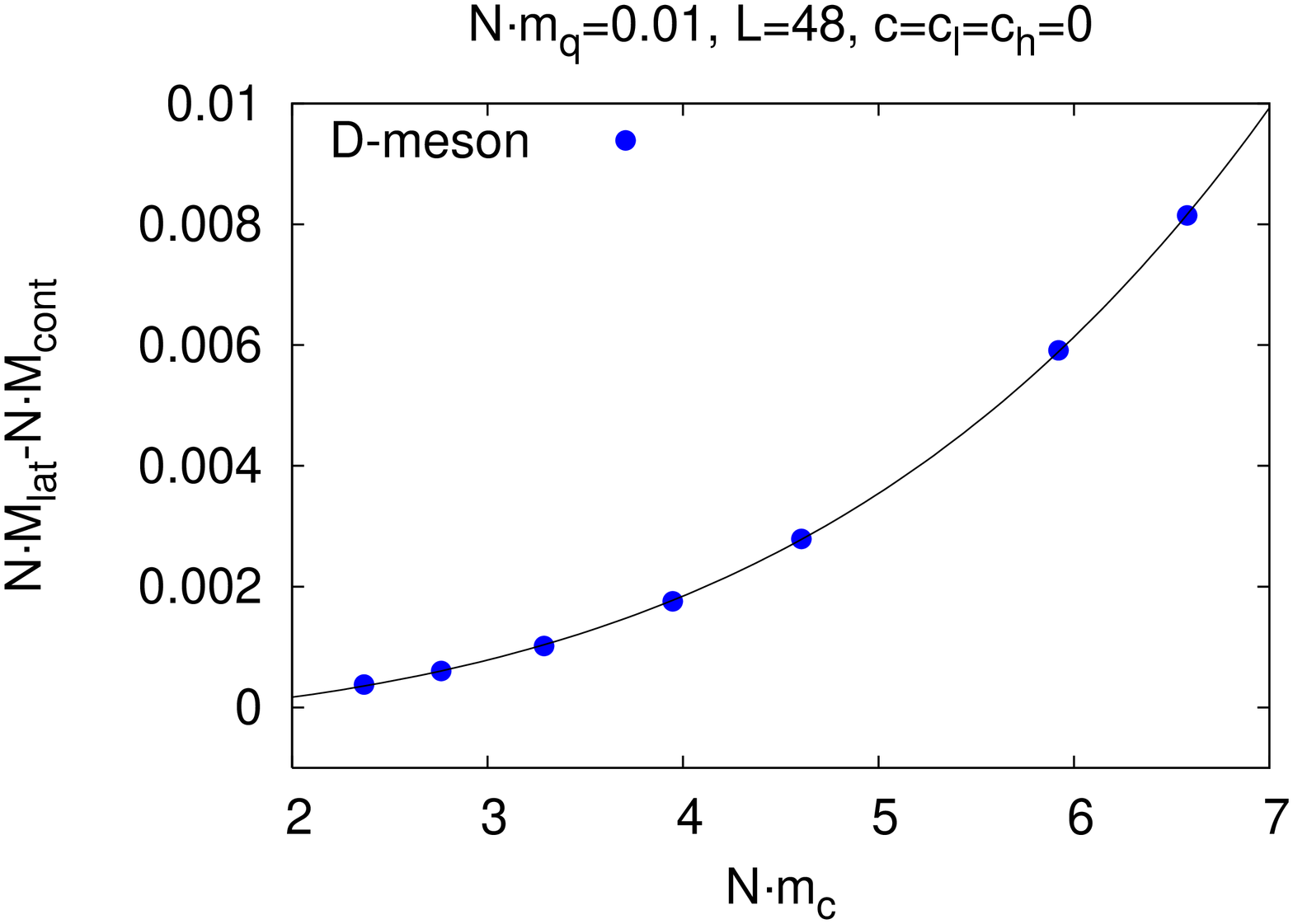}}
\caption{\label{fig:Nmq_effects} The $\mathcal{O}(a^2)$, $\mathcal{O}(a^2 m_q^2)$ and $\mathcal{O}(a^2 m_q^4)$ effects (where $q=s,c$) in: (a) kaon mass, (b) D meson mass. The case of optimal tuning to maximal twist ($c_l=c_h=0$).}
\end{figure}

In order to disentangle $\mathcal{O}(a^2)$ and $\mathcal{O}(a^2 m_q^2)$ effects, we have computed the dependence of the K and D meson masses on $Nm_s$ and $Nm_c$, respectively, for fixed lattice size $N$. Then, we have fitted the following function to the lattice data:
\begin{equation}
\label{fit-b0b1b2}
 NM_{\rm lat}-NM_{\rm cont} = b_0 \frac{1}{N^2} + b_1 \frac{(Nm_q)^2}{N^2} + b_2 \frac{(Nm_q)^4}{N^2},
\end{equation} 
where: $NM_{\rm lat}$ -- the lattice value of the K or D meson mass, $NM_{\rm cont}=Nm_{u,d}+Nm_q$ -- its continuum ($N\rightarrow\infty$) counterpart, $q=s{\,\rm or\,}c$.
The dependences of the K and D meson masses on $Nm_q$, together with the fit of the above functional form, are depicted in Fig.~\ref{fig:Nmq_effects} for the case $c_l=c_h=0$. The extracted values of $b_0,\,b_1$ and $b_2$ are shown in Table~\ref{tab:b0b1b2} for $c_l=c_h=0$, $c_l=c_h=0.1$ and $c_l=0,\,c_h=0.1$. For our fit, we have used $N=48$. However, we have checked for other lattice sizes ($N=16,\,32,\,64$) that the values of $b_0,\,b_1$ and $b_2$ remain the same, within numerical precision.

\begin{figure}[t]
\centering
\subfigure[\label{fig:b0b1a}]{\includegraphics[height=3.1in, angle=-90]{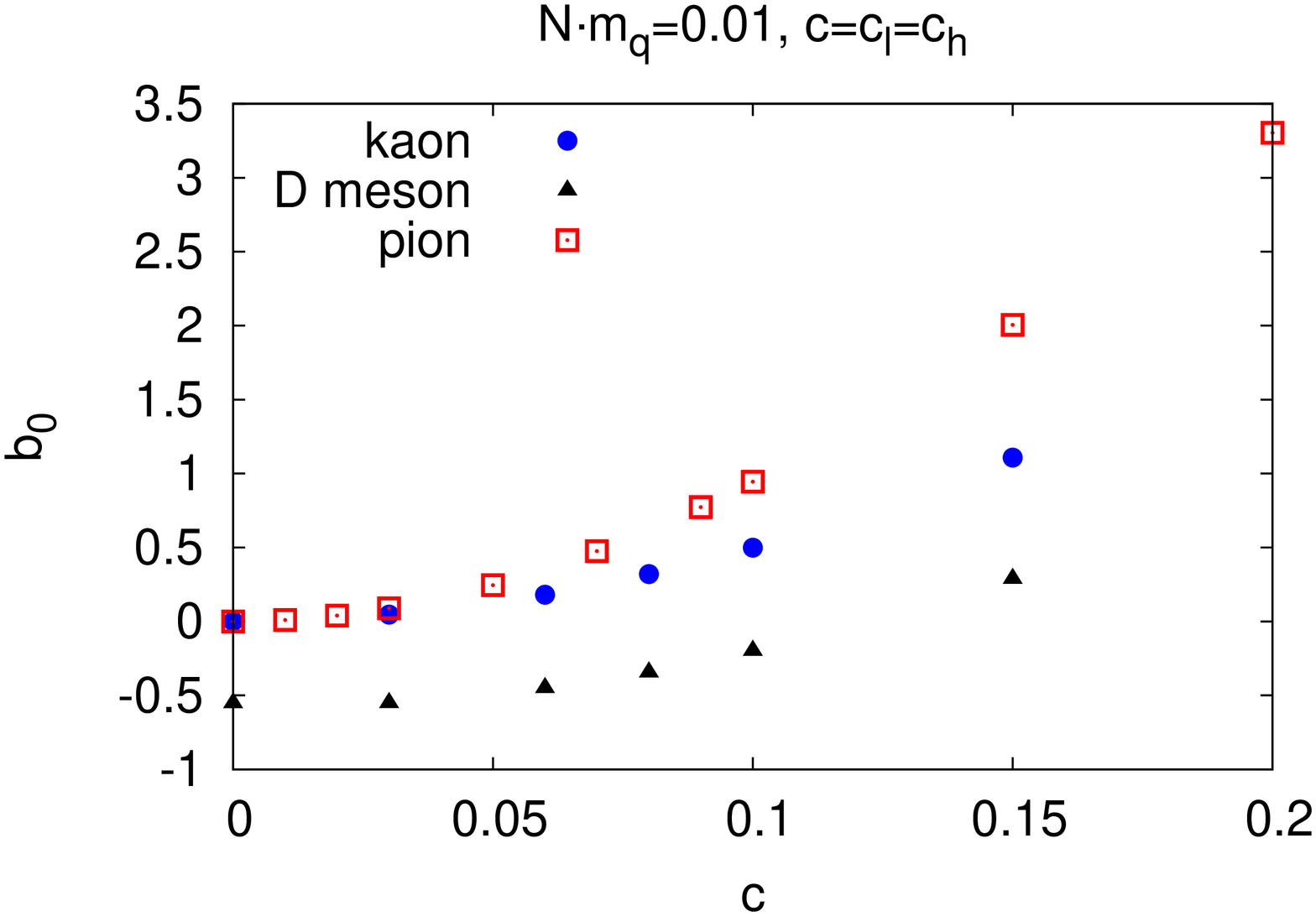}}
\subfigure[\label{fig:b0b1b}]{\includegraphics[height=3.1in, angle=-90]{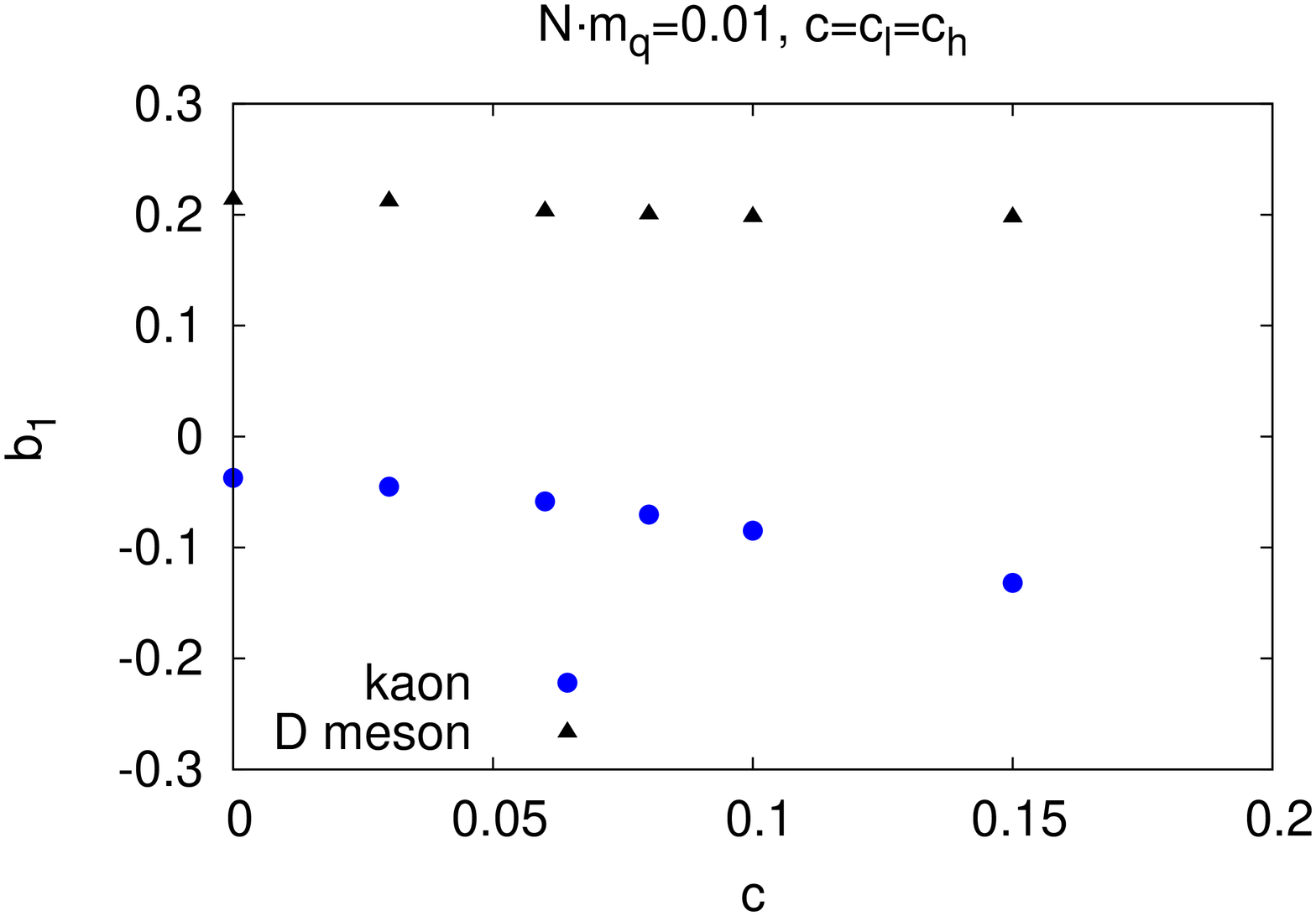}}
\caption{\label{fig:b0b1} The extracted values of the parameters (a) $b_0$, (b) $b_1$, for different values of the parameter $c=c_l=c_h$. $b_0$ and $b_1$ are defined by eq.~(\ref{fit-b0b1b2}).}
\end{figure}

\begin{figure}[t]
\centering
\subfigure[\label{fig:b0b1-cl0a}]{\includegraphics[height=3.1in, angle=-90]{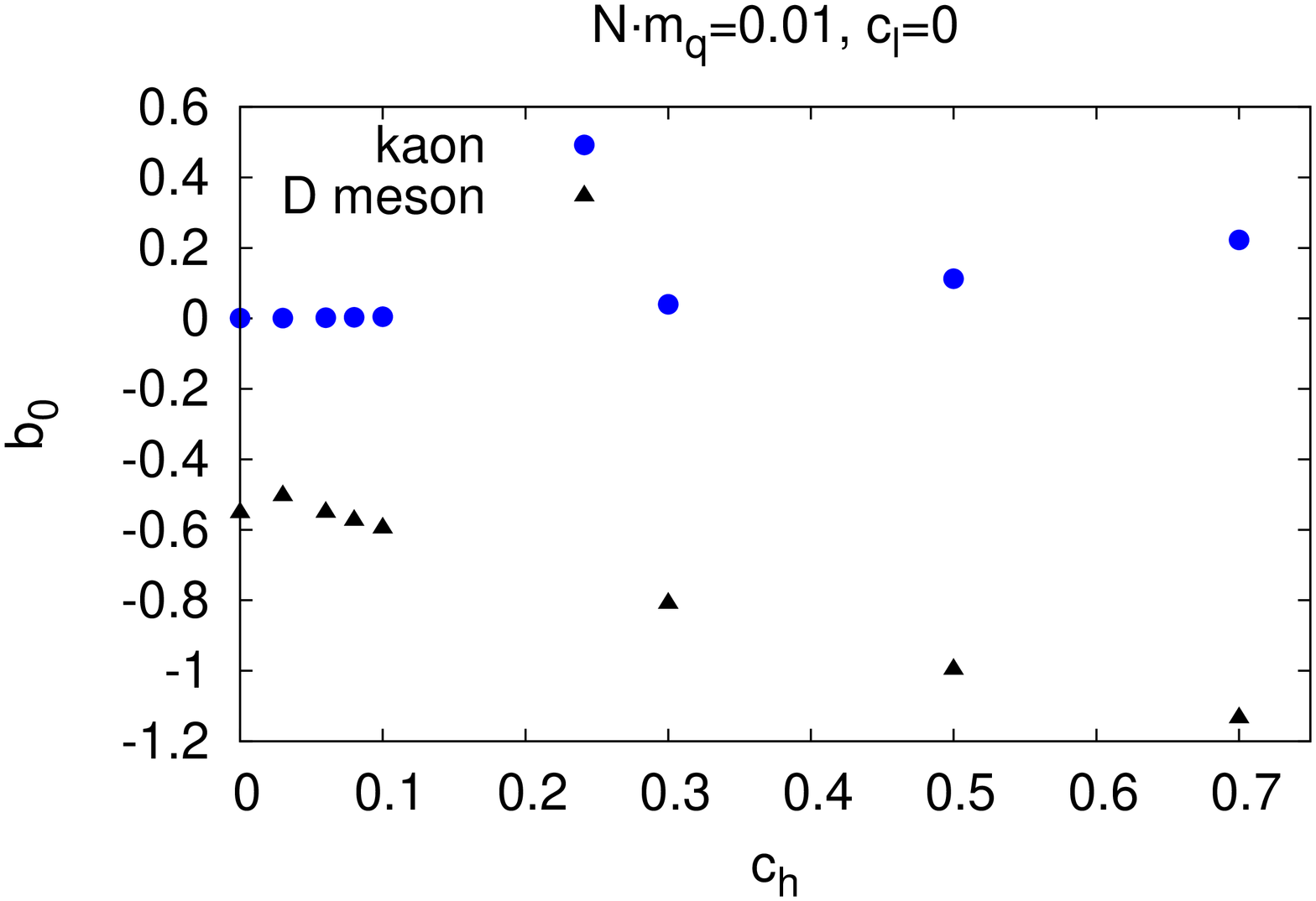}}
\subfigure[\label{fig:b0b1-cl0b}]{\includegraphics[height=3.1in, angle=-90]{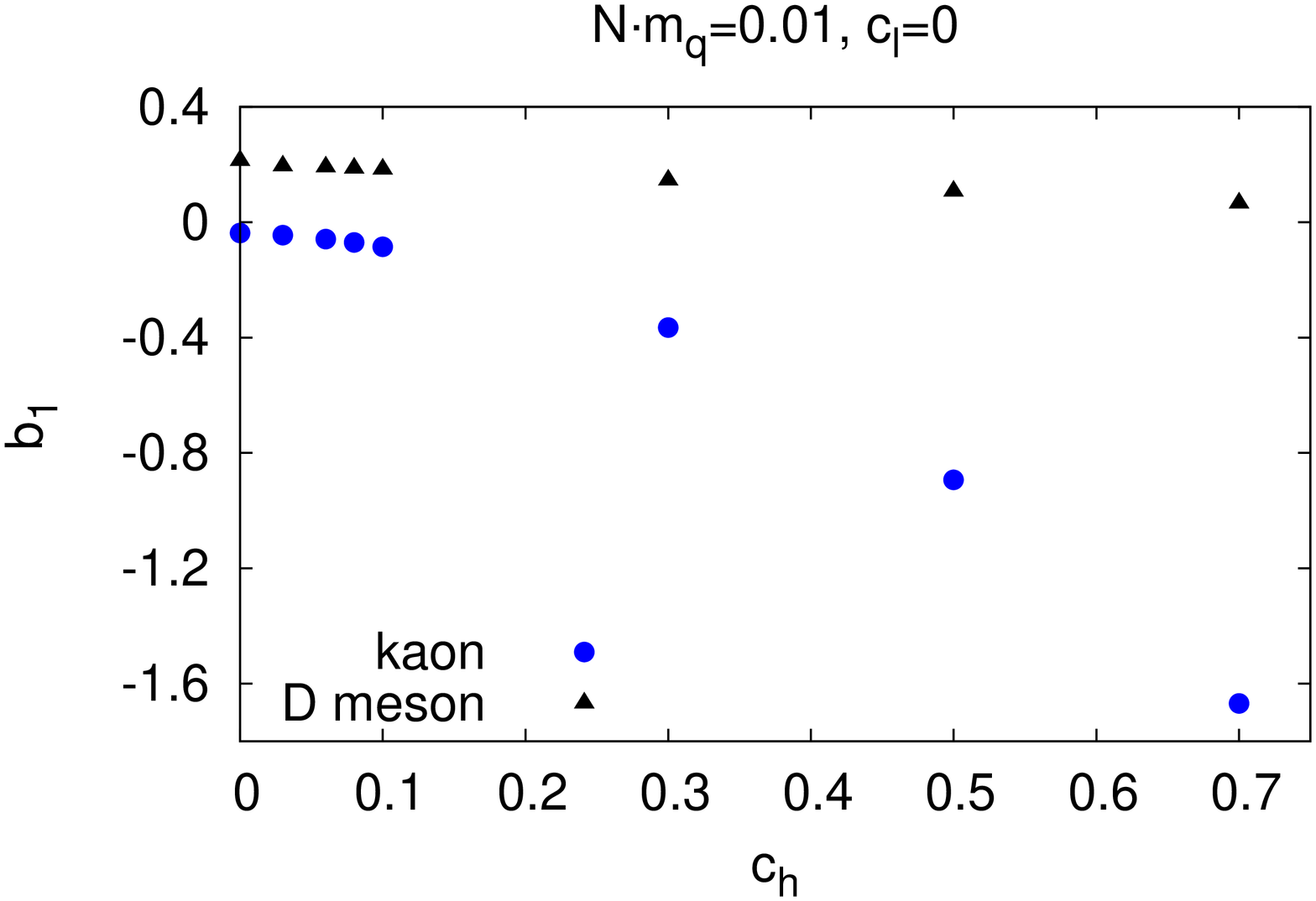}}
\caption{\label{fig:b0b1-cl0} The extracted values of the parameters (a) $b_0$, (b) $b_1$, for different values of the parameter $c_h$, with $c_l=0$. $b_0$ and $b_1$ are defined by eq.~(\ref{fit-b0b1b2}).}
\end{figure}

\begin{table}[t]
\begin{center}

\begin{tabular}{|c||c|c||c|c||c|c|}
\hline
%
 & \multicolumn{2}{c||}{$c=c_l = c_h = 0$} & \multicolumn{2}{c||}{$c=c_l = c_h = 0.1$} & \multicolumn{2}{c|}{$c_l = 0,\,c_h = 0.1$}\\
 & \multicolumn{2}{c||}{} & \multicolumn{2}{c||}{\vspace{-0.40cm}} & \multicolumn{2}{c|}{}\\

\hline
Coefficient & K meson & D meson & K meson & D meson & K meson & D meson\\
\hline
$b_0$ & 0.000571 & -0.552 & 0.499 & -0.199 & 0.00442 & -0.596\\
$b_1$ & -0.0373 & 0.214 & -0.0848 & 0.198 & -0.0848 & 0.183\\
$b_2$ & -0.159 & 0.00540 & -0.0623 & 0.00550 & -0.0623 & 0.00592\\
\hline
\end{tabular}
\end{center}
\vspace{0.2cm}
\caption{\label{tab:b0b1b2} The fitting coefficients of eq.~(\ref{fit-b0b1b2}). The fits are shown in Fig.~\ref{fig:Nmq_effects} for the case $N=48$, $c_l=c_h=0$.}

\end{table}

\begin{table}[t]
\begin{footnotesize}
\begin{center}
\begin{tabular}{|c||c|c||c|c||c|c|}
\hline
%
 & \multicolumn{2}{c||}{$c=c_l = c_h = 0$} & \multicolumn{2}{c||}{$c=c_l = c_h = 0.1$} & \multicolumn{2}{c|}{$c_l = 0,\, c_h = 0.1$}\\
 & \multicolumn{2}{c||}{} & \multicolumn{2}{c|}{\vspace{-0.4cm}} & \multicolumn{2}{c|}{}\\

\hline
Effects & K meson & D meson & K meson & D meson & K meson & D meson\\
\hline
$\mathcal{O}(a^2)$ & $2.5\cdot10^{-7}$ & $-2.4\cdot10^{-4}$ & $2164\cdot10^{-7}$ & $-0.9\cdot10^{-4}$ & $19.2\cdot10^{-7}$ & $-2.6\cdot10^{-4}$\\
$\mathcal{O}(a^2m_q^2)$ & $-7.6\cdot10^{-7}$ & $7.5\cdot10^{-4}$ & $-17\cdot10^{-7}$ & $6.9\cdot10^{-4}$ & $-17.2\cdot10^{-7}$ & $6.4\cdot10^{-4}$\\
$\mathcal{O}(a^2m_q^4)$ & $-1.5\cdot10^{-7}$ & $1.5\cdot10^{-4}$ & $-0.6\cdot10^{-7}$ & $1.6\cdot10^{-4}$ & $-0.6\cdot10^{-7}$ & $1.7\cdot10^{-4}$\\
\hline
sum & $-6.6\cdot10^{-7}$ & $6.6\cdot10^{-4}$ & $2146\cdot10^{-7}$ & $7.6\cdot10^{-4}$ & $1.4\cdot10^{-7}$ & $5.5\cdot10^{-4}$ \\
\hline
\end{tabular}
\end{center}
\vspace{0.2cm}
\caption{\label{tab:decomposition} The decomposition of the difference $NM_{\rm lat}-NM_{\rm cont}$ for the K and D mesons. The lattice size $N=48$, the strange quark mass $Nm_s=0.21591$ and the charm quark mass $Nm_c=2.84091$.}
\end{footnotesize}
\end{table}

Our fits are summarized in Fig.~\ref{fig:b0b1} (for the case $c\equiv c_l=c_h$) and Fig.~\ref{fig:b0b1-cl0} (for $c_l=0,\,c_h\geq0$).
Fig.~\ref{fig:b0b1a} shows the magnitude of ``pure'' $\mathcal{O}(a^2)$ effects for the pion, the kaon and the D meson. In the region of small values of the parameter $c$, these effects are the largest for the D meson and comparable to each other for the pion and the kaon. Moreover, the curve $b_0(c)$ coincides with the curve $a_1(c)$ in Fig.~\ref{fig:slope1}, since in the case of the pion $\mathcal{O}(a^2m_{u,d}^2)$ effects are negligible, due to the small value of light quark masses. We also observe that the value of $b_0$ increases with increasing $c$. However, since $b_0<0$ for the D meson in this region, the size of $\mathcal{O}(a^2)$ effects decreases when the parameter $c$ increases, which means that the effects of non-optimal tuning can partially cancel $\mathcal{O}(a^2)$ effects. For a particular value of the parameter $c$ (around 0.13) a complete cancellation can even occur (i.e $b_0=0$ can result).

Fig.~\ref{fig:b0b1b} shows the magnitude of $\mathcal{O}(a^2m_q^2)$ effects for the kaon and the D meson. These effects depend only slightly on the value of the parameter $c$. Interestingly, the sign of $\mathcal{O}(a^2m_q^2)$ is opposite to the sign of the ``pure'' $\mathcal{O}(a^2)$ effects, which means that a partial cancellation between these two types of effects occurs.

Further insight into the role of a method of tuning to maximal twist can be obtained by analyzing the dependence of cut-off effects for $c_l=0$ and $c_h\geq0$, i.e. if the Wilson mass $m_{0,l}$ is precisely tuned to maximal twist, while the Wilson mass $m_{0,h}$ is non-optimally tuned (Fig.~\ref{fig:b0b1-cl0}). In such case, the $\mathcal{O}(a^2)$ effects are again much larger for the D meson than for the kaon. What is more, the magnitude of these effects is much smaller for the kaon than in the case $c_l=c_h$. This implies that non-optimal tuning in the light sector leads to much larger effects than non-optimal tuning in the heavy sector. Still, both effects tend to increase the value of the coefficient $b_0$. The situation is different for the D meson. There, the effects of non-optimal tuning of $m_{0,l}$ and $m_{0,h}$ have opposite signs and thus the ``pure'' $\mathcal{O}(a^2)$ effects in the case $c_l=0,\,c_h>0$ increase with increasing value of the parameter $c_h$. In this way, the partial or complete cancellation of $\mathcal{O}(a^2)$ effects that we have observed in the case $c_l=c_h$ can be attributed to non-optimal tuning of $m_{0,l}$ and not $m_{0,h}$.

Regarding the size of $\mathcal{O}(a^2m_q^2)$ effects in the case $c_l=0,\,c_h\geq0$, Fig.~\ref{fig:b0b1-cl0b} shows that again non-optimal tuning in the light and in the heavy sector can have opposite effects. In the case of the D meson, a partial cancellation of $\mathcal{O}(a^2m_q^2)$ effects can occur for positive values of the parameter $c_h$, but in the case of the kaon, the effects of non-optimal tuning of the Wilson mass $m_{0,h}$ reinforce the magnitude of $\mathcal{O}(a^2m_q^2)$ effects.

It is also interesting to see the relative contribution of $\mathcal{O}(a^2)$, $\mathcal{O}(a^2m_q^2)$ and $\mathcal{O}(a^2m_q^4)$ discretization effects to the difference $NM_{\rm lat}-NM_{\rm cont}$ for some chosen values of the lattice size and the strange and charm quark masses. Here we choose again $N=48$ and the quark masses considered in the earlier tests described in this section, i.e. $Nm_s=0.21591$, $Nm_c=2.84091$. The decomposition of the difference $NM_{\rm lat}-NM_{\rm cont}$ for such parameter values is shown in Table \ref{tab:decomposition}, for the cases $c_l=c_h=0$, $c_l=c_h=0.1$ and $c_l=0,\,c_h=0.1$. 

The size of $\mathcal{O}(a^2)$, $\mathcal{O}(a^2m_q^2)$ and $\mathcal{O}(a^2m_q^4)$ cut-off effects is of the same order of magnitude in the case of optimal tuning to maximal twist in both the light and the heavy sector. However, in this case the largest contribution to the difference $NM_{\rm lat}-NM_{\rm cont}$ is the one of $\mathcal{O}(a^2m_q^2)$ effects, for both the K and D meson. Moreover, the relative sign of $\mathcal{O}(a^2)$ and $\mathcal{O}(a^2m_q^2)$ effects is different. The $\mathcal{O}(a^2m_q^4)$ effects are roughly a factor of 5 smaller than $\mathcal{O}(a^2m_q^2)$ effects in the case of both meson masses. Again, the overall size of discretization effects is much more important for the D meson than for the kaon, both if we consider absolute and relative cut-off effects.

In the case of non-optimal tuning of both $m_{0,l}$ and $m_{0,h}$, the $\mathcal{O}(a^2)$ effects (induced by non-optimal tuning particularly in the light sector) are much more important than $\mathcal{O}(a^2m_q^2)$ effects in the kaon case, but the latter still dominate in the case of the D meson.

An interesting situation occurs in the case $c_l=0,\,c_h=0.1$ for the kaon, where we observe that $\mathcal{O}(a^2)$ effects are almost equal in magnitude, but of opposite sign to the $\mathcal{O}(a^2m_q^2)$ effects. Thus, an almost exact cancellation occurs and the overall size of cut-off effects is much smaller than in the case of optimal tuning in both sectors. Such decrease of the overall size of discretization effects with respect to the optimal tuning case occurs also for the D meson, which is the effect that can be clearly observed in Fig.~\ref{fig:slope2}.

To summarize, we observe a very intricate interplay of different types of effects -- $\mathcal{O}(a^2)$, $\mathcal{O}(a^2m_q^2)$ and even $\mathcal{O}(a^2m_q^4)$ effects can become sizable. A partial or complete cancellation between $\mathcal{O}(a^2)$ and $\mathcal{O}(a^2m_q^2)$ effects is possible and moreover such cancellation can also occur between effects of non-optimal tuning to maximal twist in the light and heavy sectors. Needless to say, the behaviour in the interacting theory should be expected to be even more complex.

\section{Conclusions}

In this paper, we have provided an analytical basis for studying lattice spacing effects at tree-level of perturbation theory for maximally twisted mass Wilson quarks, when both the heavy quark doublet and the light one are included.
Particularly, we have calculated the 4-dimensional momentum 
space and time-momentum frame quark propagators in the heavy sector and constructed the matrix of correlation functions for the K and D mesons.

We have investigated the scaling of the kaon and 
the D meson masses with the lattice spacing,
for optimal and non-optimal tuning to maximal twist.
We have clearly verified that the lattice spacing effects
appear in even powers of $1/N=a$, as expected from the general automatic $\mathcal{O}(a)$-improvement 
of maximally twisted mass lattice QCD.
More precisely, for the case of optimal tuning (given by the introduced parameter $c\equiv c_l=c_h=0$, where $c_l$ and $c_h$ determine the light and heavy Wilson mass: $N m_{0,l} = c_l/N,\,  N m_{0,h} = c_h/N$) and 
non-optimal tuning with small $c$ values,
we have observed the linear dependence of meson masses in $1/N^2=a^2$, while for sufficiently large $c$ values,
the ${\mathcal O}(a^4)$ and higher order corrections can become important.


We have also disentangled the $\mathcal{O}(a^2)$, $\mathcal{O}(a^2m_q^2)$ and $\mathcal{O}(a^2m_q^4)$ discretization effects (where $m_q$ is the mass of the strange or the charm quark) in the kaon and the D meson masses and we have found a multitude of competing effects. The overall size of discretization effects can considerably depend on the details of tuning to maximal twist in both the light and the heavy sector. Partial cancellations (or even full cancellations for some particular values of parameters) can occur between the $\mathcal{O}(a^2)$ effects and the $\mathcal{O}(a^2m_q^2)$ effects and even between effects of non-optimal tuning in the light sector and the effects of non-optimal tuning in the heavy sector. Thus, non-optimal tuning can in certain instances decrease the overall size of discretization effects. 

We believe that although our results do not provide any proofs for the interacting case,
they give a clear warning:
one should expect a very complex interplay of different effects and thus when physical observables in the strange and especially charm sectors are evaluated in full QCD, the enhanced cut-off effects
should carefully be taken into account and physical results obtained at a single value of the lattice spacing can differ significantly from the continuum results. 


\section*{Acknowledgments}

The authors are grateful to Karl Jansen and Marc Wagner for
interesting and useful discussions.
This work was partially
supported by the DFG Sonderforschungsbereich / Transregio SFB/TR-9.
E.\,L. was supported by grants RFBR No. 09-02-00629-a, 08-02-00661-a, 09-02-00338-a, a grant for scientific schools No. NSh-679.2008.2 and
by Heisenberg Landau Program JINR-Germany collaboration.
K.\,C. was supported by Ministry of
Science and Higher Education grant nr. N N202 237437.

\section*{Appendix A}

The kernel of the heavy quark propagator:
\eq K(x;y)=-\frac{1}{2a} \sum^4_{\mu=1} \left[(\delta_{\alpha \beta}-(\gamma_{\mu})_{\alpha \beta})\delta_{x+a\hat{\mu},y}+ (\delta_{\alpha \beta}+(\gamma_{\mu})_{\alpha \beta})\delta_{x-a\hat{\mu},y}\right]\delta_{AB} \delta_{ij}+  \en
\eq \qquad\quad\;\,+\left[\left(m_{0,h}+\frac{4}{a}  \right)\delta_{\alpha\beta}\delta_{ij} + i\mu_{\sigma} (\gamma_5)_{\alpha\beta}\tau^1_{ij}+\mu_{\delta}\tau^3_{ij}\delta_{\alpha\beta}  \right]\delta_{xy} \delta_{AB},  \nonumber \en
\eq x=na, \ \ \ y=ma, \ \ \  n,m \in \mathbbm{Z}. \nonumber \en
Further, we skip colour, flavour and Lorentz indices. 

After discrete Fourier transformation: 
\eq
K(p)=\sum_{x-y}  K(x;y) e^{-ip(x-y)}, 
\en
we get:
\eq K(p)=\frac{i}{a}\sum^4_{i=1}\gamma_i \sin(p_ia)+ \frac{1}{a}\left(1-\cos(p_4 a)\right) + 
 \frac{2}{a}\sum^3_{i=1} \sin^2\left(\frac{p_i a}{2}\right) + m_{0,h} + i\mu_{\sigma} \gamma_5 \tau_1 + \mu_{\delta} \tau_3. \en
Denote:
\eq M_{0,h}=m_{0,h}+\frac{2}{a}\sum^3_{i=1}\sin^2\left( \frac{p_ia}{2} \right), \ \ \ N=M_{0,h}(p)+\frac{1}{a}(1-\cos(p_4a)), \en
\eq {\mathcal K}=\frac{1}{a}\sum^3_{i=1}\gamma_i\sin(p_i a), \ \ R^2=\mu_{\delta}^2+\mu_{\sigma}^2+{\mathcal K}^2 +\frac{1}{a^2} \sin^2(p_4 a).\en

Then, the matrix operator $K(p)$ takes the form:
\eq K(p)=N+\mu_{\delta} \tau_3 + i {\mathcal K}+i\mu_{\sigma} \gamma_5 \tau_1. \en
The propagator can be calculated from the following relation: 
\eq S^{(h)}(p)K(p)=1,  \ \ \ S^{(h)}(p)=\frac{K^{\dagger}(p)}{K(p) K^{\dagger}(p)}.  \en

Using the properties of gamma matrices, we get the heavy twisted mass propagator:
\eq S^{(h)}(p)=\frac{N+\mu_{\delta} \tau_3 - i \sum^4_{\mu=1} \frac{\gamma_{\mu}}{a} \sin(p_{\mu}a)-i\mu_{\sigma} \gamma_5 \tau_1}
{N^2+R^2+2N\mu_{\delta}\tau_3+2\mu_{\sigma}\mu_{\delta}\gamma_5 \tau_2}.\en
The denominator has the form $n+m \tau_3 + l \tau_2$, where $n,m$ and $l$ are numbers.
To eliminate the flavour structure from the denominator, we rearrange:
\eq (n+m\tau_3+l\tau_2)(n-m\tau_3-l\tau_2)=(n+m\tau_3)(n-m\tau_3)+l\tau_2(n-m\tau_3)-(n+m\tau_3)l\tau_2-l^2= \nonumber \en
\eq n^2+mn\tau_3 -nm\tau_3 -m^2+ln\tau_2-lm\tau_2 \tau_3 -nl\tau_2 - ml\tau_3 \tau_2 -l^2=n^2-m^2-l^2.\en
After this, we obtain:
\eq S^{(h)}(p))=\frac{(N+\mu_{\delta} \tau_3 - i \sum^4_{\mu=1} \frac{\gamma_{\mu}}{a} \sin(p_{\mu}a)-i\mu_{\sigma} \gamma_5 \tau_1)(N^2+R^2-2N\mu_{\delta}\tau_3-2\mu_{\sigma}\mu_{\delta}\gamma_5 \tau_2)}{(N^2+R^2)^2-4N^2 \mu_{\delta}^2-4 \mu_{\sigma}^2 \mu_{\delta}^2}.\en
The denominator of the last expression has two zeros which correspond to the poles of the propagator:
\eq \cosh E_{1,2}=\cos(p_4 a)_{1,2}=\frac{-b1 \mp \sqrt{ (b1)^2 - 4 (a1)( c1) } }{2 (a1)}.  \en

We calculate the residues of the function:
\eq f=\frac{i a}{8 \pi} e^{-E\frac{t}{a}} \frac{L1(E)}{\left(\left(M_{0,h}+\frac{1}{a}\right)^2-\mu^2_{\delta}\right)(\cosh E - \cosh E_1 )(\cosh E - \cosh E_2) } \en

The first residue is taken  at the point $E=E_1$:
\eq {\bf{(res\ f)_{E=E_1}} }=\frac{i a}{8 \pi} e^{-E_1 \frac{t}{a}} \frac{P_1(E_1)}{\left(\left(M_{0,h}+\frac{1}{a}\right)^2-\mu^2_{\delta}\right) \sinh E_1 (\cosh E_1 - \cosh E_2) },   \en
where: 
\eq P_1(E_1)=N_1 (N^2_1+R^2_1-2\mu^2_{\delta})+\mu_{\delta}(R^2_1-N_1^2-2\mu^2_{\sigma})\tau_3-i(N_1^2+R_1^2)\left({\mathcal K}+\frac{i\gamma_4}{a} \sinh E_1\right)+    \nonumber \en
\eq 2i\mu_{\sigma} \mu_{\delta} \left({\mathcal K}+\frac{i \gamma_4}{a} \sinh E_1\right) \gamma_5 \tau_2 + 2 i N_1 \mu_{\delta}\left({\mathcal K}+\frac{i \gamma_4}{a} \sinh E_1\right) \tau_3 - i \mu_{\sigma} (N_1^2+R_1^2-2\mu^2_{\delta}) \gamma_5 \tau_1, \en
\eq N_1=M_{0,h}+\frac{1}{a}(1-\cosh E_1), \ \ \ \ \ R^2_1={\mathcal K}^2+\mu_{\sigma}+\mu_{\delta}-\frac{1}{a^2} \sinh^2 E_1 . \en

The second residue is at the point $E=E_2$:

\eq {\bf(res\ f)_{E=E_2}} =\frac{i a}{8 \pi} e^{-E_2 \frac{t}{a}} \frac{P_1(E_2)}{\left(\left(M_{0,h}+\frac{1}{a}\right)^2-\mu^2_{\delta}\right) \sinh E_2 (\cosh E_2 - \cosh E_1) },   \en
where: 
\eq P_1(E_2)=N_2 (N^2_2+R^2_2-2\mu^2_{\delta})+\mu_{\delta}(R^2_2-N_2^2-2\mu^2_{\sigma})\tau_3-i(N_2^2+R_2^2)\left({\mathcal K}+\frac{i\gamma_4}{a} \sinh E_2\right)+   \nonumber \en
\eq 2i\mu_{\sigma} \mu_{\delta} \left({\mathcal K}+\frac{i \gamma_4}{a} \sinh E_2\right) \gamma_5 \tau_2 + 2 i N_2 \mu_{\delta}\left({\mathcal K}+\frac{i \gamma_4}{a} \sinh E_2\right) \tau_3 - i \mu_{\sigma} (N_2^2+R_2^2-2\mu^2_{\delta}) \gamma_5 \tau_1, \en
\eq N_2=M_{0,h}+\frac{1}{a}(1-\cosh E_2), \ \ \ \ \  R^2_2={\mathcal K}^2+\mu_{\sigma}+\mu_{\delta}-\frac{1}{a^2} \sinh^2 E_2 .\en

The propagator is then the sum: 
\eq {\bf S_{\infty}(\vec{p},t)}=2 \pi i \left[ (res\ f)_{E=E_1} + (res\ f)_{E=E_2} \right]. \en

\section*{Appendix B}

The elements of the correlation function matrix  are calculated for infinite time using the expressions for infinite time twisted mass propagators (\ref{EQN017}) and (\ref{EQN022}). We show the calculation of one of them. We use the Fourier transformation, the definition of the $\delta$-function and gamma algebra:

\eq C_{11}(t)=\sum_{\vec{x}} \langle (\bar{\psi}^{(u)}(x)\gamma_5 \psi^{(s)}(x))^{\dagger} (\bar{\psi}^{(u)}(0)\gamma_5 \psi^{(s)}(0)\rangle=\nonumber \en
\eq =\sum_{\vec{x}} {\rm Tr} \{\gamma_5 S^{\bar{u} u}(\vec{x},t;0,0) \gamma_5 S^{\bar{s}s}(0,0;\vec{x},t)\}=  \nonumber \en
\eq =\sum_{\vec{x}} {\rm Tr} \left(\frac{1}{L^3} \sum_{\vec{p}} e^{i \vec{p} \vec{x}} S^{\bar{u} u}(\vec{p},t) \frac{1}{L^3} \sum_{\vec{q}}
e^{-i \vec{q}\vec{x}} \gamma_5 S^{\bar{s}s}(\vec{q},-t)\gamma_5  \right)=  \label{EQN026} \en
\eq =\frac{1}{L^3} \sum_{\vec{p}} {\rm Tr} \left( \frac{1}{L^3}  \sum_{\vec{q}} \sum_{\vec{x}}  e^{i \vec{p} \vec{x}-i \vec{q}\vec{x}} S^{\bar{u} u}(\vec{p},t)   \gamma_5 S^{\bar{s}s}(\vec{q},-t)\gamma_5  \right)= \nonumber \en
\eq  =\frac{1}{L^3} \sum_{\vec{p}} {\rm Tr} \left( \frac{1}{L^3}  \sum_{\vec{q}} \delta(\vec{p}-\vec{q}) S^{\bar{u} u}(\vec{p},t)   \gamma_5 S^{\bar{s}s}(\vec{q},-t)\gamma_5  \right)=  \nonumber \en
\eq  =\frac{1}{L^3} \sum_{\vec{p}} {\rm Tr} \left(  S^{\bar{u} u}(\vec{p},t)   \gamma_5 S^{\bar{s}s}(\vec{p},-t)\gamma_5  \right).  \nonumber \en
The components of the light and heavy quark propagators can be written in the following way:
\eq S^{\bar{s}s}(\vec{p},-t)=\sum_{i=1}^3 \gamma_i f_i + \gamma_4 f_4 +\mathbbm{1} \cdot f_6, \ \ \  S^{\bar{u}u}(\vec{p},t)=\sum_{i=1}^3 \gamma_i d_i + \gamma_4 d_4 + \gamma_5 d_5 + \mathbbm{1} \cdot d_6. \label{EQN027}\en
Thus, we get:
\eq {\rm Tr} \left\{ S^{\bar{u} u}(\vec{p},t)\gamma_5 S^{\bar{s}s}(\vec{p},-t) \gamma_5 \right\}={\rm Tr}\Big\{ \sum_{i=1}^3 -d_i f_i-d_4 f_4+d_6 f_6 \Big\}. \label{EQN028} \en
We identify the corresponding coefficients $f_i$ and $d_i$, where $i=1,..,6$ in the expressions (\ref{EQN027}) and (\ref{EQN028}) and obtain:
\eq C_{11}(t)=\frac{N_c N_d}{L^3} \sum_{\vec{p}} \Big\{ aC{\mathcal K}^2 [A(N^2_1+R^2_1)+B(N^2_2+R^2_2)+2\mu_{\delta}(AN_1+BN_2)]+ \label{EQN029} \en 
\eq +\frac{C \sinh E_1^l}{a}\left[A(N_1^2+R_1^2)\sinh E_1+B(N_2^2+R_2^2)\sinh E_2 + 2\mu_{\delta} (AN_1\sinh E_1+BN_2 \sinh E_2)\right]+ \nonumber \en
\eq  + C(1-\cosh E_1^l+aM_{0,l})[ AN_1(N_1^2+R_1^2-2\mu^2_{\delta} )+BN_2(N_2^2+R_2^2-2\mu^2_{\delta}) - \nonumber \en
\eq  -\mu_{\delta}[A(R^2_1-N_1^2-2\mu^2_{\delta})+B(R^2_2-N^2_2-2\mu^2_{\sigma})]]\Big\}. \nonumber \en

\eq C_{22}(t)=\sum_{\vec{x}} \langle (\bar{\psi}^{(u)}(x)\gamma_5 \psi^{(c)}(x))^{\dagger} \bar{\psi}^{(u)}(0)\gamma_5 \psi^{(c)}(0)\rangle=  \nonumber \en
\eq \frac{N_c N_d}{L^3} \sum_{\vec{p}} \Big\{ aC{\mathcal K}^2 [A(N^2_1+R^2_1)+B(N^2_2+R^2_2)-2\mu_{\delta}(AN_1+BN_2)]+   \en
\eq  +\frac{C \sinh E_1^l}{a}\left[A(N_1^2+R_1^2)\sinh E_1+B(N_2^2+R_2^2)\sinh E_2 - 2\mu_{\delta} (AN_1\sinh E_1+BN_2 \sinh E_2)\right]+ \nonumber\en
\eq  +C(1-\cosh E_1^l+aM_{0,l})[ AN_1(N_1^2+R_1^2-2\mu^2_{\delta} )+BN_2(N_2^2+R_2^2-2\mu^2_{\delta}) +  \nonumber \en
\eq +\mu_{\delta}[A(R^2_1-N_1^2-2\mu^2_{\delta})+B(R^2_2-N^2_2-2\mu^2_{\sigma})]]\Big\}, \nonumber \en

\eq  C_{33}(t)=\sum_{\vec{x}} \langle (\bar{\psi}^{(u)}(x) \psi^{(s)}(x))^{\dagger} \bar{\psi}^{(u)}(0) \psi^{(s)}(0)\rangle= \nonumber \en
\eq =\frac{N_c N_d}{L^3} \sum_{\vec{p}} \Big\{ aC{\mathcal K}^2 [A(N^2_1+R^2_1)+B(N^2_2+R^2_2)+2\mu_{\delta}(AN_1+BN_2)]+ \en
\eq +\frac{C \sinh E_1^l}{a} \left[A(N_1^2+R_1^2)\sinh E_1+B(N_2^2+R_2^2)\sinh E_2 + 2\mu_{\delta} (AN_1\sinh E_1+BN_2 \sinh E_2)\right]- \nonumber \en
\eq - C(1-\cosh E_1^l+aM_{0,l})[ AN_1(N_1^2+R_1^2-2\mu^2_{\delta} )+BN_2(N_2^2+R_2^2-2\mu^2_{\delta}) -  \nonumber \en
\eq - \mu_{\delta}[A(R^2_1-N_1^2-2\mu^2_{\delta})+B(R^2_2-N^2_2-2\mu^2_{\sigma})]]\Big\}, \nonumber \en

\eq C_{44}(t)=\sum_{\vec{x}} \langle (\bar{\psi}^{(u)}(x) \psi^{(c)}(x))^{\dagger} \bar{\psi}^{(u)}(0) \psi^{(c)}(0)\rangle= \nonumber \en
\eq =\frac{N_c N_d}{L^3} \sum_{\vec{p}} \Big\{ aC{\mathcal K}^2 [A(N^2_1+R^2_1)+B(N^2_2+R^2_2)-2\mu_{\delta}(AN_1+BN_2)]+ \en
\eq +\frac{C\sinh E_1^l}{a} \left[A(N_1^2+R_1^2)\sinh E_1+B(N_2^2+R_2^2)\sinh E_2 - 2\mu_{\delta} (AN_1\sinh E_1+BN_2 \sinh E_2)\right]- \nonumber\en
\eq -C(1-\cosh E_1^l+aM_{0,l})[ AN_1(N_1^2+R_1^2-2\mu^2_{\delta} )+BN_2(N_2^2+R_2^2-2\mu^2_{\delta})+ \nonumber \en
\eq +\mu_{\delta}[A(R^2_1-N_1^2-2\mu^2_{\delta})+B(R^2_2-N^2_2-2\mu^2_{\sigma})]]\Big\}, \nonumber \en

\eq C_{43}(t)=\sum_{\vec{x}} \langle (\bar{\psi}^{(u)}(x) \psi^{(s)}(x))^{\dagger} \bar{\psi}^{(u)}(0) \psi^{(s)}(0)\rangle= \nonumber \en
\eq =\frac{N_c N_d}{L^3} \sum_{\vec{p}} aC \mu_{q} \mu_{\sigma} \{A(N_1^2+R^2_1-2\mu^2_{\delta})+B(N^2_2+R^2_2-2\mu^2_{\delta}) \}, \en 

\eq C_{34}(t)=\sum_{\vec{x}} \langle (\bar{\psi}^{(u)}(x) \psi^{(c)}(x))^{\dagger} \bar{\psi}^{(u)}(0) c(0)\rangle= \nonumber \en
\eq =\frac{N_c N_d}{L^3} \sum_{\vec{p}} aC \mu_{q} \mu_{\sigma} \{A(N_1^2+R^2_1-2\mu^2_{\delta})+B(N^2_2+R^2_2-2\mu^2_{\delta}) \}, \en 

\eq C_{31}(t)=\sum_{\vec{x}} \langle (\bar{\psi}^{(u)}(x) \psi^{(s)}(x))^{\dagger} \bar{\psi}^{(u)}(0) \gamma_5 \psi^{(s)}(0)\rangle= \nonumber \en
\eq =\frac{N_c N_d}{L^3} \sum_{\vec{p}} (iaC\mu_q)\{  AN_1(N^2_1+R^2_1-2\mu^2_{\delta})+BN_2(N^2_2+R^2_2-2\mu^2_{\delta})- \en
\eq -\mu_{\delta} [A(R^2_1-N^2_1-2\mu^2_{\sigma})+B(R^2_2-N^2_2-2\mu^2_{\sigma})]\}, \nonumber \en

\eq C_{13}(t)=\sum_{\vec{x}} \langle (\bar{\psi}^{(u)}(x) \gamma_5 \psi^{(s)}(x))^{\dagger} \bar{u}(0)  \psi^{(s)}(0)\rangle= \nonumber \en
\eq =\frac{N_c N_d}{L^3} \sum_{\vec{p}} (-iaC\mu_q)\{  AN_1(N^2_1+R^2_1-2\mu^2_{\delta})+BN_2(N^2_2+R^2_2-2\mu^2_{\delta})- \en
\eq -\mu_{\delta} [A(R^2_1-N^2_1-2\mu^2_{\sigma})+B(R^2_2-N^2_2-2\mu^2_{\sigma})]\}, \nonumber \en

\eq C_{42}(t)=\sum_{\vec{x}} \langle (\bar{\psi}^{(u)}(x) \psi^{(c)}(x))^{\dagger} \bar{\psi}^{(u)}(0) \gamma_5 \psi^{(c)}(0)\rangle= \nonumber \en
\eq =\frac{N_c N_d}{L^3} \sum_{\vec{p}} (iaC\mu_q) \{  AN_1(N^2_1+R^2_1-2\mu^2_{\delta})+BN_2(N^2_2+R^2_2-2\mu^2_{\delta})+ \en
\eq +\mu_{\delta} [A(R^2_1-N^2_1-2\mu^2_{\sigma})+B(R^2_2-N^2_2-2\mu^2_{\sigma})]\},  \nonumber \en

\eq C_{24}(t)=\sum_{\vec{x}} \langle (\bar{\psi}^{(u)}(x) \gamma_5 \psi^{(c)}(x))^{\dagger} \bar{\psi}^{(u)}(0)  \psi^{(c)}(0)\rangle= \nonumber \en
\eq =\frac{N_c N_d}{L^3} \sum_{\vec{p}} (-iaC\mu_q) \{  AN_1(N^2_1+R^2_1-2\mu^2_{\delta})+BN_2(N^2_2+R^2_2-2\mu^2_{\delta})+ \en
\eq +\mu_{\delta} [A(R^2_1-N^2_1-2\mu^2_{\sigma})+B(R^2_2-N^2_2-2\mu^2_{\sigma})]\}, \nonumber \en

\eq C_{12}(t)=\sum_{\vec{x}} \langle (\bar{\psi}^{(u)}(x) \gamma_5 \psi^{(s)}(x))^{\dagger} \bar{\psi}^{(u)}(0) \gamma_5  \psi^{(c)}(0)\rangle= \nonumber \en
\eq =\frac{N_c N_d}{L^3} \sum_{\vec{p}} (-aC\mu_q \mu_{\sigma}) \left\{A(N^2_1+R^2_1-2\mu^2_{\delta})+B(N^2_2+R^2_2-2\mu^2_{\delta})\right\},\en

\eq C_{21}(t)=\sum_{\vec{x}} \langle (\bar{\psi}^{(u)}(x) \gamma_5 \psi^{(c)}(x))^{\dagger} \bar{\psi}^{(u)}(0) \gamma_5  \psi^{(s)}(0)\rangle= \nonumber \en
\eq =\frac{N_c N_d}{L^3} \sum_{\vec{p}} (-aC\mu_q \mu_{\sigma}) \left\{A(N^2_1+R^2_1-2\mu^2_{\delta})+B(N^2_2+R^2_2-2\mu^2_{\delta})\right\},\en

\eq C_{14}(t)=\sum_{\vec{x}} \langle (\bar{\psi}^{(u)}(x) \gamma_5 \psi^{(s)}(x))^{\dagger} \bar{\psi}^{(u)}(0)  \psi^{(c)}(0)\rangle= \nonumber \en
\eq = \frac{N_c N_d}{L^3} \sum_{\vec{p}} \{ -2iCa{\mathcal K}^2 \mu_{\sigma} \mu_{\delta} (A+B)-
\frac{2iC}{a} \sinh E_1^l \mu_{\sigma}\mu_{\delta}(A \sinh E_1+B \sinh E_2)-   \en
\eq -iC\mu_{\sigma}(1-\cosh E_1^l+aM_{0,l})[A(N^2_1+R^2_1-2\mu^2_{\delta})+B(N^2_2+R^2_2-2\mu^2_{\delta})]\}, \nonumber \en

\eq C_{41}(t)=\sum_{\vec{x}} \langle (\bar{\psi}^{(u)}(x) \psi^{(c)}(x))^{\dagger} \bar{\psi}^{(u)}(0) \gamma_5 \psi^{(s)}(0)\rangle= \nonumber \en
\eq =\frac{N_c N_d}{L^3} \sum_{\vec{p}} \{ 2iCa{\mathcal K}^2 \mu_{\sigma} \mu_{\delta} (A+B)+
\frac{2iC}{a} \sinh E_1^l \mu_{\sigma}\mu_{\delta}(A \sinh E_1+B \sinh E_2)+  \en
\eq +iC\mu_{\sigma}(1-\cosh E_1^l+aM_{0,l})[A(N^2_1+R^2_1-2\mu^2_{\delta})+B(N^2_2+R^2_2-2\mu^2_{\delta})]\}, \nonumber \en

\eq C_{32}(t)=\sum_{\vec{x}} \langle (\bar{\psi}^{(u)}(x) \psi^{(s)}(x))^{\dagger} \bar{\psi}^{(u)}(0) \gamma_5 \psi^{(c)}(0)\rangle=  \nonumber \en
\eq =\frac{N_c N_d}{L^3} \sum_{\vec{p}} \{ -2iCa{\mathcal K}^2 \mu_{\sigma} \mu_{\delta} (A+B) 
-\frac{2iC}{a} \sinh E_1^l \mu_{\sigma} \mu_{\delta} (A \sinh E_1+ B \sinh E_2)+ \en
\eq +iC\mu_{\sigma}(1-\cosh E_1^l+a M_{0,l})[A(N^2_1+R^2_1-2\mu^2_{\delta})+B(N^2_2+R^2_2-2\mu^2_{\delta})]\}  \nonumber \en

\eq C_{23}(t)=\sum_{\vec{x}} \langle (\bar{\psi}^{(u)}(x) \gamma_5  \psi^{(c)}(x))^{\dagger} \bar{\psi}^{(u)}(0)  \psi^{(s)}(0)\rangle= \nonumber \en
\eq =\frac{N_c N_d}{L^3} \sum_{\vec{p}} \{ 2iCa{\mathcal K}^2 \mu_{\sigma} \mu_{\delta} (A+B) 
+\frac{2iC}{a} \sinh E_1^l \mu_{\sigma} \mu_{\delta} (A \sinh E_1+ B \sinh E_2)- \en
\eq -iC\mu_{\sigma}(1-\cosh E_1^l+a M_{0,l})[A(N^2_1+R^2_1-2\mu^2_{\delta})+B(N^2_2+R^2_2-2\mu^2_{\delta})]\}.  \nonumber \en

\section*{Appendix C}

From the $4\times 4$ correlation matrix in the twisted basis presented in App.B we get the masses of K and D mesons by solving a generalized eigenvalue problem \cite{Blossier:2009kd}:
\eq 
\sum_k C_{jk} (t) {v}_k^{(n)} (t,t_0)= \sum_k C_{jk}(t_0) {v}_k^{(n)} (t,t_0) \lambda^{(n)}(t,t_0),
\label{eq:egvl_prob}
\en
where $k$  runs over $(h, \Gamma)$, \ $h \in \{s,c\}$ and $\Gamma \in \{\gamma_5, 1  \}$, $n=0,...,3$.

Eq. (\ref{eq:egvl_prob}) leads to the following equation, giving the four effective masses  $m^{(n)}_{\rm eff}$  for the lattice with temporal extension $T$:
 \eq
 \frac{\lambda^{(n)} (t,t_0)}{\lambda^{(n)} (t+1,t_0)}=
\frac{e^{{-m^{(n)}_{\rm eff}}(t,t_0) t} + e^{{-m^{(n)}_{\rm eff}}(t,t_0) (T-t)}}{e^{{-m^{(n)}_{\rm eff}}(t,t_0) (t+1)} + e^{{-m^{(n)}_{\rm eff}}(t,t_0) (T-(t+1))}}.
\label{eq:eqvl_equation}
 \en
  
Since we have performed the integration over $p_4$, our masses correspond to the limit of infinite $T$. We consider finite $t$ values, so the formula (\ref{eq:eqvl_equation}) becomes a ratio of two exponentials. 
For our tree level calculations, we have used $t_0 \geqslant 10 $. We have explored the dependence of the K and D masses on time and we have found that for the considered spatial extensions $L$ the mass plateaus are reached at $t\lesssim 100$.

We extract the meson masses from the plateaus at $t \in [50,120]$ (depending on the spatial extension $L$ of the lattices). In this range of $t$, our numerical precision gives plateaus of a very good quality and thus reliable data for the masses. 
  
To determine the parity and flavour content of the four effective masses  we perform the approximate rotation to  pseudo physical basis  \cite{Baron:2010th}. After the  rotation, the correlation matrix takes the form: 
\eq C^{\rm ph}_{(h,\Gamma)}(t)= M(\omega_l, \omega_h) C_{(h,\Gamma)}^{\chi}(t) M(\omega_l, \omega_h)^{\dagger}. \label{EQN031}\en
 The  twist rotation matrix is orthogonal at maximal twist and has the form:
 \eq M(\omega_l, \omega_h) = \left( \begin{matrix} 
\cos{\frac{\omega_l}{2}} \cos{\frac{\omega_h}{2}}  & -\sin{\frac{\omega_l}{2}} \sin{\frac{\omega_h}{2}} 
  & -i \sin{\frac{\omega_l}{2}} \cos{\frac{\omega_h}{2}}  & -i \cos{\frac{\omega_l}{2}} \sin{\frac{\omega_h}{2}} \\
-\sin{\frac{\omega_l}{2}} \sin{\frac{\omega_h}{2}}  & \cos{\frac{\omega_l}{2}} \cos{\frac{\omega_h}{2}}
  & -i \cos{\frac{\omega_l}{2}} \sin{\frac{\omega_h}{2}}  & -i \sin{\frac{\omega_l}{2}} \cos{\frac{\omega_h}{2}} \\
-i\sin{\frac{\omega_l}{2}} \cos{\frac{\omega_h}{2}}  & -i\cos{\frac{\omega_l}{2}} \sin{\frac{\omega_h}{2}}
  &  \cos{\frac{\omega_l}{2}} \cos{\frac{\omega_h}{2}}  & -\sin{\frac{\omega_l}{2}} \sin{\frac{\omega_h}{2}} \\
-i\cos{\frac{\omega_l}{2}} \sin{\frac{\omega_h}{2}}  & -i\sin{\frac{\omega_l}{2}} \cos{\frac{\omega_h}{2}}
  & - \sin{\frac{\omega_l}{2}} \sin{\frac{\omega_h}{2}}  &  \cos{\frac{\omega_l}{2}} \cos{\frac{\omega_h}{2}} 
\end{matrix} \right) \label{EQN032}\en
and connects the  operator ${\mathcal O}^{\rm ph}_{(h,\Gamma)}$ in pseudo  physical basis with the ${\mathcal O}^{\chi}_{(h,\Gamma)}$ in the twisted mass one: 
\eq {\mathcal O}^{\rm ph}_{(h,\Gamma)} = M(\omega_l, \omega_h) {\mathcal O}^{\chi}_{(h,\Gamma)},
 \ \ \  ({\mathcal O}^{\rm ph}_{(h,\Gamma)})^{\dagger} = ({\mathcal O}^{\chi}_{(h,\Gamma)})^{\dagger} M^{T}(\omega_l, \omega_h).  \label{EQN033} \en
These bilinear operators are chosen in the following way:
\eq
 {\mathcal O}^{\rm ph}_{(h,\Gamma)} = \left( \begin{matrix} 
 \bar{\psi}^{(u)} \gamma_5 \psi^{s} \\
 \bar{\psi}^{(u)} \gamma_5 \psi^{c} \\
 \bar{\psi}^{(u)} \psi^{s} \\
 \bar{\psi}^{(u)} \psi^{c} 
\end{matrix} \right), \ \ \ \ \ \ \ \ 
{\mathcal O}^{\chi}_{(h,\Gamma)} = \left( \begin{matrix} 
 \bar{\chi}^{(u)} \gamma_5 \chi^{s} \\
 \bar{\chi}^{(u)} \gamma_5 \chi^{c} \\
 \bar{\chi}^{(u)} \chi^{s} \\
 \bar{\chi}^{(u)} \chi^{c} 
\end{matrix} \right). 
\label{EQN034} \en


\end{document}